\documentclass[fleqn,usenatbib,useAMS]{mnras}
\usepackage{multirow}
\usepackage{graphicx}
\usepackage{xcolor}
\usepackage{hyperref}
\usepackage{comment}
\usepackage{soul}
\usepackage[nolist]{acronym}
\usepackage{booktabs}
\usepackage{longtable}
\usepackage[utf8]{inputenc}
\usepackage{lineno}

\begin{acronym}
    \acro{ecdf}[ECDF]{Empirical Cumulative Distribution Function}
    \acro{gls}[GLS]{Generalized Lomb Scargle}
    \acro{pdf}[PDF]{Probability Density Function}
    \acro{psd}[PSD]{Power Spectral Density}
\end{acronym}

\title[The variability patterns of PG 1553+113]{The variability patterns of the TeV blazar PG 1553+113 from a decade of MAGIC and multi-band observations}
\author[H.~Abe~et.~al.]{\parbox{\textwidth}{\Large{
MAGIC Collaboration: H.~Abe$^{1}$,
S.~Abe$^{1}$,
J.~Abhir$^{2}$,
V.~A.~Acciari$^{3}$,
I.~Agudo$^{4}$,
T.~Aniello$^{5}$,
S.~Ansoldi$^{6,42}$,
L.~A.~Antonelli$^{5}$,
A.~Arbet Engels$^{7}$,
C.~Arcaro$^{8}$
% ...
B.~Patricelli$^{16,5}$,  
G.~Ghirlanda$^{76,79}$, 
R.~Lico$^{78,4}$ \\ 
\emph{\normalsize Affiliations are listed at the end of the paper}
}}}

\newcommand{\pg}{PG$~$1553+113~}
\newcommand{\pgn}{PG$~$1553+113}

\newcommand{\fermilat}{\emph{Fermi}-LAT}
\date{Accepted XXX. Received YYY; in original form ZZZ}

\pubyear{2023}

\begin{document}
\label{firstpage}
\pagerange{\pageref{firstpage}--\pageref{lastpage}}
\maketitle

\clearpage

\begin{abstract}
{PG~1553+113 is one of the few blazars with a convincing quasi-periodic emission in the gamma-ray band. The source is also a very high-energy (VHE; $>$100\,GeV) gamma-ray emitter. \\
To better understand its properties and identify the underlying physical processes driving its variability, the MAGIC Collaboration initiated a multiyear, multiwavelength monitoring campaign in 2015 involving the OVRO 40-m and Medicina radio telescopes, REM, KVA, and the MAGIC telescopes, \textit{Swift} and \textit{Fermi} satellites, and the WEBT network. The analysis presented in this paper uses data until 2017 and focuses on the characterization of the variability. 
The gamma-ray data show a (hint of a) periodic signal compatible with literature, but the X-ray and VHE gamma-ray data do not show statistical evidence for a periodic signal. In other bands, the data are compatible with the gamma-ray period, but with a relatively high p-value. 
The complex connection between the low and high-energy emission and the non-monochromatic modulation and changes in flux suggests that a simple one-zone model is unable to explain all the variability. Instead, a model including a periodic component along with multiple emission zones is required.
}
\end{abstract}


\section{Introduction}
Variability on a wide range of time scales is a common trait of active galactic nuclei \citep[AGN, see ][for a recent review]{2019NewAR..8701541H}.
Among AGN, the blazar subclass is dominated by the emission from the relativistic outflow emerging from the supermassive black hole (SMBH) and structured in beamed jets, namely, forming a small angle with the observer. The relativistic beaming amplifies the observed radiation and results in a broad spectral energy distribution (SED) dominated by the non-thermal continuum, showing two main humps. The first peaks in the infrared (IR) to X-ray region, presumably originated by synchrotron radiation, and the second hump spans from the UV up to the TeV. 
Blazars are distinguishable by their flux variability, with a large amplitude -- up to two orders of magnitude-- and fast variations, down to a few minutes. Depending on the equivalent width (EW) of the absoption line in the optical band, blazars are further divided into Flat Spectrum Radio Quasars (FSRQs, EW $>$ 5\,\AA) and BL Lac objects (BL Lacs, EW $<$ 5\,\AA). 
The location of the synchrotron peak in blazars determines a further division into low- ($\nu^S_{peak}<10^{14}$\,Hz), intermediate- ($10^{14}<\nu^S_{peak}<10^{15}$\,Hz), and high-synchrotron peak ($\nu^S_{peak}>10^{15}$\,Hz) BL Lacs.
The mechanisms driving the variability in blazars are still debated, and several different interpretations were suggested, related to the physical processes in the jet or in the accretion mechanism ~\citep[e.g.,][and references therein]{Raiteri_CTA102,Marscher_2016galaxies}.

Jet emission models can be tested by looking at the variability pattern and correlations between the low-energy and high-energy humps of the SED \citep{2014A&ARv..22...73F,2019ApJ...880...32L}, built through contemporaneous multiwavelength (MWL) monitoring campaigns \citep{1997ARA&A..35..445U}. 

In this framework,  the discovery of regular and periodic patterns in otherwise apparently stochastic variability can provide a deeper insight into the underlying processes. 

\pg is one of the brightest HBL emitting at gamma-ray energies. The search of intervening absorption in its far-UV spectrum suggest a redshift in the range 0.413 $<$ z $<$  0.56 \citep{2016ApJ...817..111D,2010ApJ...720..976D}. Recent optical/UV observations seems to converge on a value of 0.433 for the redshift of this source \citep{2022MNRAS.509.4330D,2019ApJ...884L..31J}.

Its comparably large redshift entails a strong attenuation of the gamma-ray flux above energies E $>$ 250\,GeV due to pair production with photons of the extragalactic background light (EBL). Albeit this attenuation, \pg is a well known TeV gamma-ray emitter observed by all major imaging air Cherenkov telescopes (IACTs) \citep{2015ApJ...802...65A,2012ApJ...748...46A,2015ApJ...799....7A}.
 
The continuous gamma-ray lightcurve was measured with the Large Area Telescope (LAT) on board the \textit{Fermi} Gamma-ray Space Telescope since 2008 and shows a clear signature of a periodic modulation of $2.18\pm 0.08$ years  at E\,$>100$\,MeV and E\,$>1$\,GeV, covering 3.5 cycles\,\citep{2015ApJ...813L..41A}. The periodicity has $< 1\%$ chance of being due to random variability. This was the first time such a periodicity has been found convincingly in a gamma-ray blazar. The signature has been confirmed by several other works even in recent times \citep[e.g.][]{2020ApJ...895..122C,2020ApJ...896..134P}. 
Optical fluxes correlate with gamma-ray emission at 99\% confidence and the optical lightcurve shows evidence for modulation of $2.06 \pm 0.05$ years over 4.5 cycles. 

The mechanisms underlying the daily/weekly variability typical of blazars and the claimed periodicity are not fully understood.
The long-term monitoring of the variable lightcurves of such objects is a powerful tool for discovering such processes, and observations with a complete MWL coverage are needed to understand them.

The last years have witnessed the growth of new key observations of possible periodic behaviours in AGN and blazars, evaluated with different methods \citet[e.g.][]{2017ApJ...847....7B,2020A&A...634A.120A}, and also \citet{2019MNRAS.482.1270C} for a cautious approach. The source \pg has been used extensively to this end \citep[e.g.,][]{2017MNRAS.465..161S}. 
The periodicity should be compared to the lifetime and duty cycle of AGN activity, generally assumed as $10^7-10^8$ years, or to episodic nuclear activity and jet formation $\sim 10^5$ years \citep{1993MNRAS.263..168H,2013MNRAS.430.2137K}. For this reason, any claim of periodicity must establish compelling statistical evidence emerging from stochastic fluctuations that can mimic a periodic pattern. 

Because the emission of HBLs across the electromagnetic spectrum is dominated by the jet, the quasi-periodic modulation is most probably associated with the jet itself or with the processes at its base. In the latter case, disk perturbations or instabilities can induce a variation of the accretion rate advected to the jet, with quasi-periodic behaviour  \citep[e.g.,][]{Tchekhovskoy:2011qv}.
Among the possible interpretations, the modulation can be driven by the interaction of two SMBHs \citep[see e.g.,][and references therein]{2015ApJ...813L..41A}. Therefore, \pg is a candidate for harboring a close binary SMBH system with milli-pc separation in the early inspiral gravitational-wave driven regime prior to coalescence \citep[see e.g.,][]{2018ApJ...854...11T}. However, different viable solutions are possible. The observed modulation can be related to the jet itself, such as jet precession or intrinsically rotating flow or helical jet, or to the process feeding the jet.
Each interpretation may lead to different expectations about the evolution of the modulation at different wavelengths. 
Geometrical models, e.g., due to jet precession \citep{2000A&A...360...57R}, rotation \citep{1992A&A...255...59C,2000AA...359..948R}, or helical structure \citep{1993ApJ...411...89C}, would produce a quasi-periodic variation of the beaming factor due to the change of the viewing angle \citep{2004ApJ...615L...5R}. Therefore, almost achromatic variability is expected at all wavelengths, or with clear correlations between different wavelengths \citep{1999A&A...347...30V}.
To test this hypothesis, \pg was monitored with the very long baseline array (VLBA) at 15, 24, and 43 GHz in the period 2015-2017, a full cycle of gamma-ray activity \citep{2020A&A...634A..87L}. 
VLBA data provided evidence of jet angle variations, indicating that geometric effects could play a role in the observed emission variability through Doppler boosting modulation. However, no clear connection was found between the observed variations and the quasi-periodic variability patterns reported in optical and gamma rays. Therefore, additional mechanisms are necessary to explain the variable broadband emission. 


Pulsating instabilities occurring in the disk can explain the quasi-periodicity, also in the case of a single SMBH \citep{1992PASJ...44..529H}, as in magnetically dominated and magnetically arrested accretion flows in the inner disc part \citep{2011MNRAS.418L..79T}.

In case the observed periodicity is interpreted as a periodic perturbation due to the interaction with a secondary black hole, a double or multiple peak substructure is expected at different bands, resulting from the interaction of the secondary black hole with the accretion disk near the periastron \citep{1996ApJ...460..207L}. Such a structure is apparent in the optical lightcurve, and the detection of similar double peaks in the X-ray and gamma-ray lightcurves would confirm this interpretation.

 \pg is also known to be variable on a weekly time scale \citep{2015MNRAS.450.4399A,2015ApJ...802...65A,2015ApJ...799....7A}.  The study of flaring episodes provided an important input for the modelling of intrinsic emission from the source and was used as a probe of fundamental physics \citep{2020arXiv200207571G} and of the EBL \citep{2019MNRAS.486.4233A,2020A&A...633A..74K}.
Finally, the reported periodicity could be accidental due to stochastic background fluctuations, typically found in lightcurves of AGN and blazars. For this reason, proper statistical approaches such as those discussed in \citet{2003MNRAS.345.1271V} are needed when studying long-term light curves \citep[e.g.,][]{2019MNRAS.482.1270C}. Multiwavelength observations are needed to support the physical interpretation of the emission from the innermost regions of the blazar and its jet.

With the purpose of characterizing \pg  broadband variability  and testing  physical scenarios, the MAGIC collaboration initiated a multi-year, MWL monitoring campaign in 2015. In the campaign, various instruments observing in the radio, infrared, optical (both photometry and polarimetry), UV, soft X-ray, and gamma-ray bands were involved. In this article we report the results of this campaign, including data from previous observations.
The paper is structured as follows: in \S 2, we present the details of the MAGIC and MWL data analyses. \S~3 is dedicated to the characterization of the source variability, while \S~4 is focused on periodicity study of the data presented in the paper. In \S~5 the conclusions of this work are outlined.

\section{Multiwavelenth data and analysis}
The MWL data used in this study span several orders of magnitude, from radio to VHE gamma rays. In this section, we give a brief description of the collected data and their analyses (from higher to lower frequencies), while in the following sections we report on the scientific interpretation of the data. 

\subsection{VHE gamma-ray data}

MAGIC is a system of two IACTs located in La Palma, Canary Islands, at 2200~m asl. It observes VHE gamma rays from 50~GeV up to tens of TeV. The angular resolution at around 200~GeV energies at low zenith angles (0$^{\circ}$-30$^{\circ}$), is $<$ 0.07$^{\circ}$, while the energy resolution is 16\% \citep{2016APh....72...76A}. From 2004 to 2009, MAGIC was composed of a single IACT, MAGIC-I \citep{2004NIMPA.518..188B}.

The MAGIC data are a collection of images registered by the camera of each telescope and are processed with a standard analysis chain \citep{2013ICRC...33.2937Z}. 

Since its detection in 2004, MAGIC observed PG~1553+113 every year. In the first two years of observation (2005 and 2006) the data were affected by large systematic and statistical errors, therefore, in this work we consider only data taken from 2007 on, where an upgrade of the telescope readout sensibly increased its performances \citep{2016APh....72...61A}. 
For the study, we use the dark night data from 2007 to 2017, including already published data from 2007, 2008, 2009, and 2012 campaigns \citet{2012ApJ...748...46A,2015MNRAS.450.4399A}. 

Table~\ref{tab:MAGICdata} summarises the results of the MAGIC data analysis of the samples collected from 2007 to 2017. 
Data from 2010 on were analysed with monthly binning, while for previous data, the overall yearly sample is considered as reported in \citet{2012ApJ...748...46A}. 
The total observation time is $\sim$100\,hours, non uniformly distributed across the years (third column). Average fluxes above 150\,GeV in physical units and in Crab Nebula units\footnote{The Crab Unit (C.U.) used in this work is an arbitrary unit obtained by dividing the integral energy flux measured above a certain threshold by the Crab Nebula flux measured above the same threshold by MAGIC as reported in \citet{2016APh....72...76A}.} are reported in columns four and five. The flux varies from 4 to 18\% C.U. across time. Finally, the last three columns list the results of the differential flux analysis, reporting the results of a fit with a simple power law function in the form
 \begin{equation}
 \frac{dF}{dE}= f_0 * {\left( \frac{E}{200\,{\rm GeV}}\right)}^{-\Gamma}
 \label{simpeq}
\end{equation}
where $f_0$ is the flux at 200\,GeV (column six) and $\Gamma$ is the power law index (column seven in Table~\ref{tab:MAGICdata}).

\begin{figure*}
\centering
\includegraphics[width=0.8\textwidth]{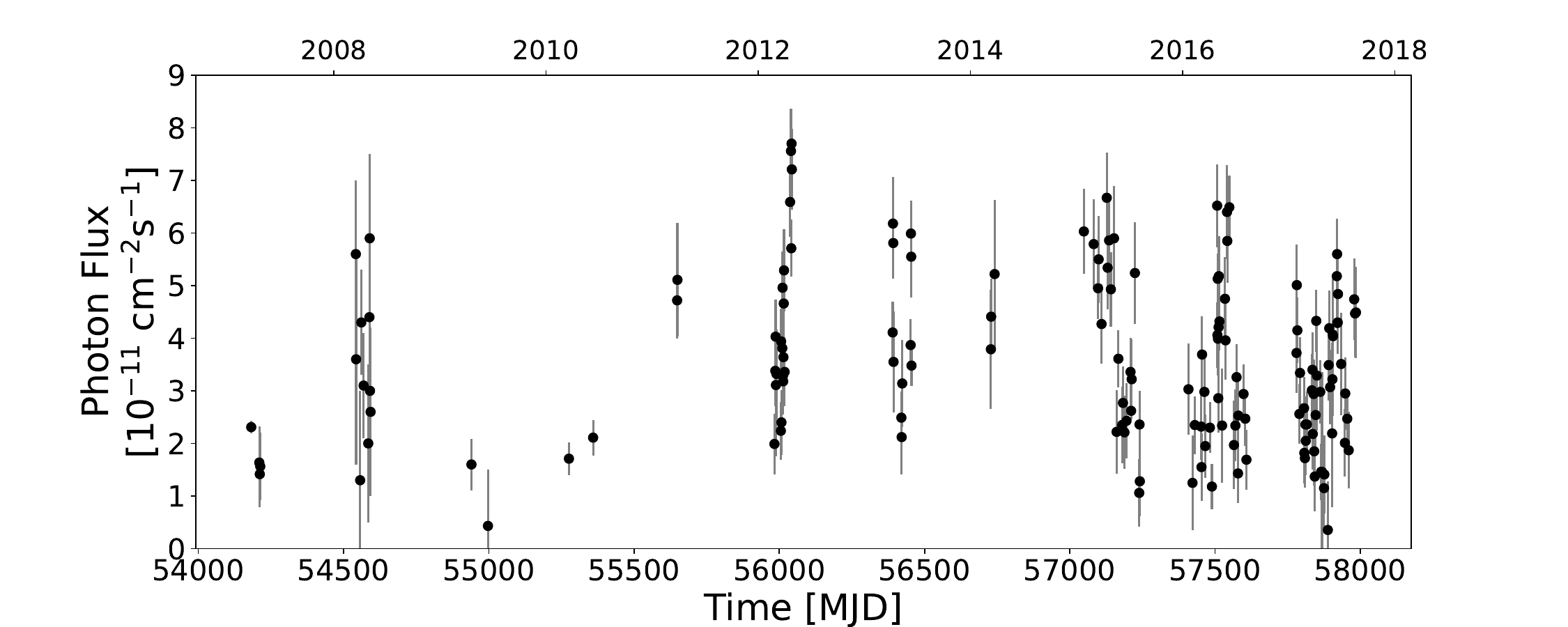}
\caption{PG~1553+113 lightcurve above 150\,GeV measured with the MAGIC telescopes.}
\label{Fig:MAGIC_LC}
\end{figure*}

\begin{table*}
\caption{Summary of MAGIC telescopes observations. }
\centering
\begin{tabular}{l|l|l|l|l|l|l|l}
\hline
\hline
Year & Month & Time & Average Flux  $>150$~GeV &  F$_{>150~GeV}$  &   f$_0$   & $\Gamma$  \\ 
     &       & [h]  & [cm$^{-2}$s$^{-1}$] & C.U.$^{a}$           & [cm$^{-2}$ s$^{-1}$ TeV$^{-1}$]   &  \\
\hline
\hline
2007$^{b}$&   & 11.5  & (1.40 $\pm$ 0.38)  $\cdot$ 10$^{-11}$   & 4\% &  (1.1 $\pm$ 0.3) $\cdot$ 10$^{-10}$  & 4.1 $\pm$ 0.3  \\
\hline
2008$^{b}$&   & 8.7   & (3.70 $\pm$ 0.47)  $\cdot$ 10$^{-11}$    & 11\% & (2.6 $\pm$ 0.3) $\cdot$ 10$^{-10}$ & 4.3 $\pm$ 0.4 \\
\hline
2009$^{b}$&   & 8.5   & (1.63 $\pm$ 0.45)  $\cdot$ 10$^{-11}$   & 5\% & (1.3 $\pm$ 0.2) $\cdot$ 10$^{-10}$ & 3.6 $\pm$ 0.5 \\ 
\hline 
2010 & March & 3.19 & (1.71 $\pm$ 0.31)  $\cdot$ 10$^{-11}$  & 5\% &(1.1 $\pm$ 0.1) $\cdot$ 10$^{-10}$ & 3.8 $\pm$ 0.7\\
     & June  & 2.85 & (2.11 $\pm$ 0.34)  $\cdot$ 10$^{-11}$  & 6\%& (1.8 $\pm$ 0.2) $\cdot$ 10$^{-10}$ & 3.2 $\pm$ 0.7  \\
\hline 
2011 & March & 2.19 &(4.77 $\pm$ 0.61)   $\cdot$ 10$^{-11}$ & 14\% & (3.2 $\pm$ 0.4) $\cdot$ 10$^{-10}$ & 3.4 $\pm$ 0.3\\
\hline
2012 & February & 1.94 & (2.54 $\pm$ 0.42)  $\cdot$ 10$^{-11}$ & 8\%&   (1.8  $\pm$ 0.3)$\cdot$ 10$^{-10}$ & 3.6 $\pm$ 0.4\\
     & March    & 11.58 & (3.77 $\pm$ 0.27)$\cdot$ 10$^{-11}$ & 11\%  & (2.4 $\pm$ 0.2)$\cdot$ 10$^{-10}$  & 3.6 $\pm$ 0.2\\
     & April    & 8.87 & (5.85 $\pm$ 0.34) $\cdot$ 10$^{-11}$  & 18\% & (3.3 $\pm$ 0.2)$\cdot$ 10$^{-10}$  & 3.7 $\pm$ 0.1\\
\hline 
2013 & April  & 4.00 & (5.13 $\pm$ 0.37) $\cdot$ 10$^{-11}$ & 16\% & (3.0 $\pm$ 0.3) $\cdot$ 10$^{-10}$ & 3.5 $\pm$ 0.2 \\ 
     & May    & 2.52 & (2.21 $\pm$ 0.39) $\cdot$ 10$^{-11}$ & 7\% & (1.7 $\pm$ 0.2) $\cdot$ 10$^{-10}$ &  3.7 $\pm$ 0.3 \\
     & June   & 6.35 & (4.22 $\pm$ 0.28) $\cdot$ 10$^{-11}$ & 13\%  & (2.6 $\pm$ 0.2) $\cdot$ 10$^{-10}$ & 3.1 $\pm$ 0.1 \\ 
\hline
2014 & March & 1.97 &(4.09 $\pm$ 0.55) $\cdot$ 10$^{-11}$    & 12\%  & (3.0 $\pm$ 0.4) $\cdot$ 10$^{-10}$ & 3.4 $\pm$ 0.4 \\
\hline
2015 & January & 1.13 & (5.98 $\pm$ 0.80) $\cdot$ 10$^{-11}$   & 18\%  & (3.9 $\pm$ 0.7) $\cdot$ 10$^{-10}$ &  4.5 $\pm$ 0.4 \\ 
     & March   & 4.72 & (5.09 $\pm$ 0.36) $\cdot$ 10$^{-11}$   & 15\%  & (3.2 $\pm$ 0.2) $\cdot$ 10$^{-10}$ &  3.9 $\pm$ 0.1\\ 
     & April   & 4.16 & (5.59 $\pm$ 0.40) $\cdot$ 10$^{-11}$   & 17\%  & (3.5 $\pm$ 0.2) $\cdot$ 10$^{-10}$ &  3.9 $\pm$ 0.1\\ 
     & May     & 3.65 & (3.83 $\pm$ 0.43) $\cdot$ 10$^{-11}$   & 12\%  & (2.3 $\pm$ 0.2) $\cdot$ 10$^{-10}$ &  3.8 $\pm$ 0.3\\ 
     & June    & 3.73 & (2.51 $\pm$ 0.35) $\cdot$ 10$^{-11}$   & 8\% & (1.8 $\pm$ 0.3) $\cdot$ 10$^{-10}$ &  3.5 $\pm$ 0.4 \\ 
     & July    & 3.64 & (4.00 $\pm$ 0.44) $\cdot$ 10$^{-11}$   & 12\% & (2.3 $\pm$ 0.3) $\cdot$ 10$^{-10}$ &  3.7 $\pm$ 0.2\\ 
     & August  & 4.47 & (1.70 $\pm$ 0.39) $\cdot$ 10$^{-11}$   & 5\%  & (1.4 $\pm$ 0.4) $\cdot$ 10$^{-10}$ &  3.9 $\pm$ 0.4\\ 
\hline
2016 & January    &0.96 & (2.76 $\pm$ 0.86) $\cdot$ 10$^{-11}$   & 8\%  &  n.a.$^{\dagger}$   & n.a.$^{\dagger}$  \\
& February    &2.35 & (2.19 $\pm$ 0.48) $\cdot$ 10$^{-11}$   & 7\% &  (1.4 $\pm$ 0.3) $\cdot$ 10$^{-10}$ & 3.8 $\pm$ 0.9 \\
& March    & 5.19 & (2.31 $\pm$ 0.30) $\cdot$ 10$^{-11}$   & 7\%  & (1.5 $\pm$ 0.2) $\cdot$ 10$^{-10}$ & 3.4 $\pm$ 0.2 \\
& April-a    & 4.16 & (3.33 $\pm$ 0.29) $\cdot$ 10$^{-11}$   & 10\% & (1.9 $\pm$ 0.2) $\cdot$ 10$^{-10}$ & 4.1 $\pm$ 0.2 \\
& April-b    & 2.28 & n.a.   & n.a. &  n.a.   & n.a.  \\
& May    & 10.37 & (4.22 $\pm$ 0.23) $\cdot$ 10$^{-11}$   & 13\%  & (2.6 $\pm$ 0.1) $\cdot$ 10$^{-10}$ & 3.6 $\pm$ 0.1 \\
& June    & 4.43 & (5.81 $\pm$ 0.38) $\cdot$ 10$^{-11}$   & 18\%  & (3.2 $\pm$ 0.2) $\cdot$ 10$^{-10}$ & 3.7 $\pm$ 0.1 \\
& July    & 5.68 & (2.57 $\pm$ 0.29) $\cdot$ 10$^{-11}$   & 8\%  & (1.7 $\pm$ 0.2) $\cdot$ 10$^{-10}$ & 3.5 $\pm$ 0.2 \\
& August    & 3.25 & (2.25 $\pm$ 0.38) $\cdot$ 10$^{-11}$   & 7\% & (1.4 $\pm$ 0.3) $\cdot$ 10$^{-10}$ & 3.2 $\pm$ 0.2 \\
\hline
2017 & January    & 3.15 & (4.17 $\pm$ 0.41) $\cdot$ 10$^{-11}$   & 13\%  & (2.4 $\pm$ 0.2) $\cdot$ 10$^{-10}$ & 3.9 $\pm$ 0.2\\
& February    & 7.74 & (2.39 $\pm$ 0.22) $\cdot$ 10$^{-11}$   & 7\%  & (1.6 $\pm$ 0.1) $\cdot$ 10$^{-10}$ & 4.1 $\pm$ 0.1\\
& March    & 6.44 & (2.46 $\pm$ 0.26) $\cdot$ 10$^{-11}$   & 7\% & (1.5 $\pm$ 0.1) $\cdot$ 10$^{-10}$ & 4.0 $\pm$ 0.1\\
& April    & 5.43 & (2.77 $\pm$ 0.27) $\cdot$ 10$^{-11}$   & 8\% & (1.7 $\pm$ 0.1) $\cdot$ 10$^{-10}$ & 3.8 $\pm$ 0.1\\
& May    & 5.67 & (2.82 $\pm$ 0.28) $\cdot$ 10$^{-11}$   & 8\% & (1.8 $\pm$ 0.1) $\cdot$ 10$^{-10}$ & 3.8 $\pm$ 0.1\\
& June    & 9.80 & (4.55 $\pm$ 0.24) $\cdot$ 10$^{-11}$   & 14\%  & (2.3 $\pm$ 0.1) $\cdot$ 10$^{-10}$ & 3.9 $\pm$ 0.1\\
& July    & 3.27 & (2.53 $\pm$ 0.34) $\cdot$ 10$^{-11}$   & 8\% & (1.7 $\pm$ 0.2) $\cdot$ 10$^{-10}$ & 3.7 $\pm$ 0.2\\
& August    & 3.0 & (4.50 $\pm$ 0.45) $\cdot$ 10$^{-11}$   & 14\% & (3.1 $\pm$ 0.3) $\cdot$ 10$^{-10}$ & 3.9 $\pm$ 0.3\\
\hline
\hline
\end{tabular}

$^{a}$ C.U. is the Crab Unit, as defined in the text. $^{b}$ Observations performed with a single telescope, MAGIC-I and published in \citet{2012ApJ...748...46A}. $^{\dagger}$ Observation was too short to allow for a reliable fit to the spectrum.
\label{tab:MAGICdata}
\end{table*}

The overall emission above 150\,GeV of PG~1553+113 observed with MAGIC from 2007 to 2017 is reported in Figure \ref{Fig:MAGIC_LC}. Data from 2007, 2008, and 2009 were collected with a single telescope and are from \citet{2012ApJ...748...46A}. For the more recent data, a daily binning was adopted.
The 2010-2017 monthly-averaged values are listed in Table~\ref{tab:MAGICdata}, along with 2007-2009 yearly values from \citet{2012ApJ...748...46A}. MAGIC started a regular monitoring of the source for seven months per year in 2015 (MJD $\sim 57000$). This explains the irregular and scarce sampling of the curve before 2015. 

The daily flux above 150\,GeV shows variations within a factor of $\sim$\,10, and reached its maximum in 2012 during a historical flare reported in \citet{2015MNRAS.450.4399A}. 
The average flux is (2.74$\pm$0.04) $\cdot$ 10$^{-11}$cm$^{-2}$s$^{-1}$. The hypothesis of constant flux can be discarded, based on the $\chi^2$ test for goodness of fit ($\chi^2$/(degrees of freedom)=1339/157).

\subsection{High energy gamma-ray data}
Further gamma-ray data considered in the study are those collected with the \fermilat\, and analysed above 100\,MeV in \citet{2018ApJ...854...11T}. 
In that work the authors added 2016 and 2017 data to the sample used in the original paper claiming the quasi-periodic behaviour \citep{2015ApJ...813L..41A}. 
The \fermilat\, data therefore cover almost continuously the 2008--2017 period and represent the only continuous monitoring considered. These data are analysed with 20 days binning.

\subsection{X-ray data}

\begin{figure*}
\centering
\includegraphics[width=0.8\textwidth]{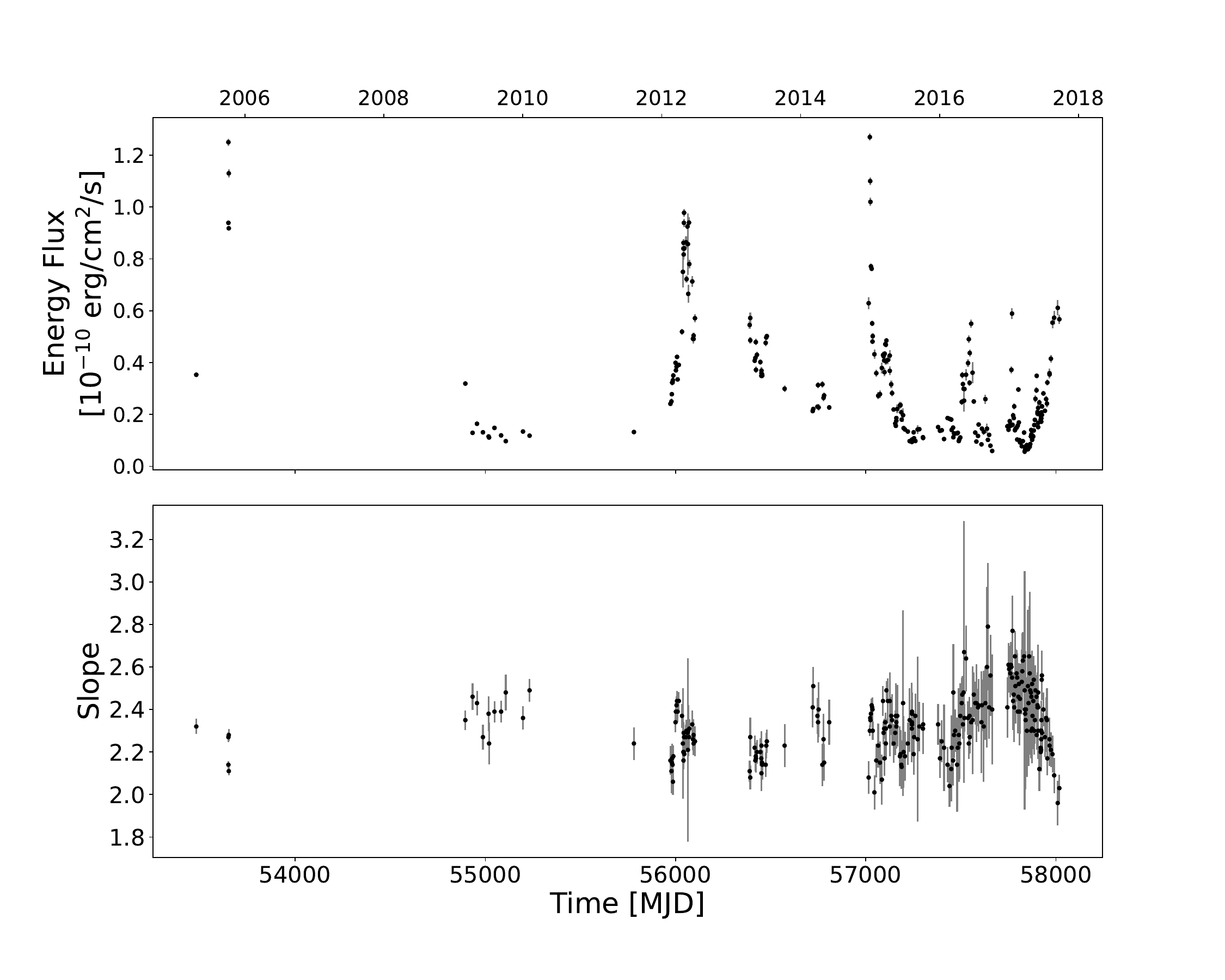}
\caption{Top panel: PG~1553+113 lightcurve in the 0.5-10\,keV band as detected with the \textit{Swift}-XRT satellite. Bottom panel: spectral slope lightcurve resulting from an absorbed power-law fit.}
\label{Fig:XRT_LC}
\end{figure*}

The \textit{Neil Gehrels Swift observatory (Swift)} \citep{2004ApJ...611.1005G} observed \pg since 2005 during outbursts and almost regularly since 2013. We have collected all snapshots in the period  from 2005 up to the end of 2017. 
\pg was observed simultaneously with the the X-ray Telescope \citep[XRT;][0.2--10.0 keV]{2005SSRv..120..165B}, and with all six filters of  the Ultraviolet/Optical Telescope \citep[UVOT;][170--600 nm]{2005SSRv..120...95R}.
The \textit{Swift}-XRT data, reported in Fig.\ref{Fig:XRT_LC}, were collected in photon counting mode (PC) and windowed timing mode (WT). In both cases, the data were processed using the FTOOLS task \texttt{xrtpipeline} (version 0.13.5), which is distributed by HEASARC within the HEASoft package (v6.28). Events with grades $0-12$ were selected for the data in PC mode, and with grades $0-2$ for the data in WT mode. The corresponding response matrices available in the {\it Swift} CALDB version were used.
When the source count rate in PC mode was higher than 0.6 counts s$^{-1}$ the  pile-up was evaluated following the standard procedure\footnote{\url{http://www.swift.ac.uk/analysis/xrt/pileup.php}}. Observations affected by pile-up were corrected by masking the central  $7$ arcsec region. For each observation the extraction region was checked visually on the image and slightly centred to the peak of the signal, when needed. The signal was extracted within an annulus with an inner radius of 3 pixels (7 arcsec) and an outer radius of 30 pixels (70 arcsec).    
Events in different channels were grouped with the corresponding redistribution matrix (rmf), and ancillary (arf) files with the task \texttt{grppha}, setting a binning of at least 25 counts for each spectral channel in order to use the chi-squared statistics. The resulting spectra were analyzed with \texttt{Xspec} version 12.11.1. We fitted the spectrum with an absorbed power-law using the photoelectric absorption model \texttt{tbabs} \citep{wilms00}, with a neutral hydrogen column density fixed to its Galactic value \citep[$N_{\rm H} = 3.67\times10^{20}$\,cm$^{-2}$;][]{2005A&A...440..775K}. The results are shown in Fig~\ref{Fig:XRT_LC}.

\subsection{UV data}

{\em Swift}/UVOT data in the $v$, $b$, $u$, $w1$, $m2$, and $w2$ filters are reported in Fig.~\ref{Fig:MWL_LC} and were reduced with the \texttt{HEAsoft} package v6.28 using the \texttt{uvotsource} task. We extracted the source counts from a circle with 5 arcsec radius centred on the source nominal position, corresponding to the optimal aperture on the source count rate \citep{2008MNRAS.383..627P}. The background counts were extracted from a circle with 60 arcsec radius in a near, source-free region. Conversion of magnitudes into dereddened flux densities was obtained by adopting the extinction value E(B--V) = 0.054 as in \citet{Raiteri:2015lr}, the mean galactic extinction curve in \citet{1999PASP..111...63F} and the magnitude-flux calibrations by \citet{Caldb02}.  
Statistical uncertainty on magnitudes of the order of 0.03 mag,  on the zero-point UVOT calibration 0.02-0.06 mag \citep{Caldb01} and the count ratio to flux correction \citep{Caldb02} have been propagated to estimate the error on the flux, resulting in a 4\% to 6\% uncertainty.

\subsection{Optical data}
The optical R-band data were obtained as part of Tuorla blazar monitoring program\footnote{\url{http://users.utu.fi/kani/1m} \citep{2008AIPC.1085..705T}}. 
The observations are described in \citet{2018A&A...620A.185N}.
The data have been analysed using the semiautomatic pipeline for differential photometry developed at the Tuorla Observatory \citep{2018A&A...620A.185N} using the comparison and control star magnitudes from \citet{2015MNRAS.454..353R}. The observed fluxes have been corrected for galactic extinction using a value of 0.113 from \citet{2011ApJ...737..103S}.

Other optical data were provided by the Whole Earth Blazar Telescope (WEBT) Collaboration\footnote{\url{https://www.oato.inaf.it/blazars/webt/} \citep{2002A&A...390..407V}.}. WEBT observations up to 2015 October were analysed in \citet{2015MNRAS.454..353R, 2017MNRAS.466.3762R}. New data in the 2016 and 2017 optical observing seasons were acquired at the following observatories: Abastumani (Georgia), Aoyama Gakuin (Japan), Crimean (Crimea\footnote{In 1991, Ukraine with the Crimean peninsula became an independent state. While the Crimean Astrophysical Observatory
became Ukrainian, the AZT-8 telescope located there continued to be operated jointly by the Crimean observatory and by the St. Petersburg group.}),  Hans Haffner (Germany), Mt.\ Maidanak (Uzbekistan), Perkins (US), Rozhen (Bulgaria), Siding Spring (Australia), Siena (Italy), Sirio (Italy), St.\,Petersburg (Russia), Teide (Spain), Tijarafe (Spain), and at the Astronomical Station Vidojevica (Serbia). Additional WEBT observations were carried out with telescopes belonging to the Las Cumbres Observatory global telescope network at the Haleakala, Siding Spring, and Teide observing sites.

Calibration was performed using the same photometric sequence as in the case of KVA data. 
The $R$-band lightcurve obtained by assembling all datasets is shown in Fig.~\ref{Fig:MWL_LC} and was carefully inspected and, when necessary, processed to obtain a homogeneous and reliable time series. Indeed, even if WEBT observers use the same photometric sequence, differences in equipment may lead to some offset between various datasets. These offsets clearly appear when datasets overlap in time and can consequently be corrected for. Moreover, we removed a few data points strongly deviating from the main trend traced by the bulk of the datasets, and mostly affected by large uncertainties. Finally, noisy intra-night sequences from the same telescope were binned. The above processing is a necessary step to undertake if one wants to deal with light curves that can be used for robust analysis and modelling.

\subsection{Optical Polarisation}
We use optical polarisation data obtained by the Nordic Optical Telescope (NOT), Liverpool Telescope (RINGO2), Skinakas Observatory (RoboPol), Crimean Astrophysical Observatory (AZT-8+ST7 telescope and LX-200 telescope with SBIG ST7b), Perkins Telescope Observatory, and Steward Observatory. The NOT data reduction is described in \citet{hovatta16,2018MNRAS.480..879M}. The RINGO2 data were obtained as a part of a blazar monitoring programme at the Liverpool Telescope \citep{jermak16} and the RoboPol data as a part of a blazar monitoring programme at the Skinakas Observatory \citep{blinov21}. The details of the data reduction of the AZT-8+ST7 data are described in  \citet{larionov08}, and the Perkins telescope observations in \cite{jorstad10}. The Steward Observatory data are publicly available and the polarimetric data are described in detail in \cite{smith09}. 

The data were obtained using the R-band filter except for the Steward Observatory where data were obtained using a filter between $5000-7000$\,\AA. All data were checked for consistency and the polarisation degree was corrected for positive bias using the formula in \cite{wardle74}. We removed six data points, which had a signal-to-noise ratio less than two in fractional polarisation. In 2016, RoboPol observed the source with a faster cadence of multiple observations per night, and we averaged these to a single observation per night to avoid biasing our {\bf analysis} with more densely sampled curves during that time. 

In Fig.~\ref{Fig:MWL_LC}, we show the EVPAs starting from a range between $0^\circ$ and $180^\circ$. The difference between the EVPAs of consecutive points is minimised to be less than $90^\circ$ by adding or subtracting $180^\circ$ from the following points. If the time gap between the points is longer than 50 days, we set the EVPA to the original range of $0^\circ-180^\circ$ as we do not know the evolution of the EVPA over such long gaps. 

In all other cases, except for the Steward Observatory observations, photometry is also determined from the observations. The R-band magnitudes from these observations are also included in Fig.~\ref{Fig:MWL_LC} along with the KVA and WEBT data. The polarisation data are used in this paper for charaterising the general variability patterns, while more detailed physical modelling of the polarisation is the subject of a separate paper (Nilsson et al. in prep.).

\subsection{IR data}
We observed \pg with the Rapid Eye Mounting Telescope \citep[REM,][]{2004SPIE.5492.1590Z}, a robotic telescope located at La Silla Observatory (Chile). It performed photometric observations using NIR filters in the period from April 08, 2005 (MJD 53468) to June 20, 2017 (MJD 57802). REM data are shown in Fig.~\ref{Fig:MWL_LC}.
The telescope is able to operate in a fully  autonomous way \citep{2004SPIE.5492.1613C}, and data are reduced and analysed following standard procedures. Aperture photometry was derived using custom tools, and the calibration was based on the scheme described by \cite{2014A&A...562A..79S}. 

We used reference stars from the \textit{Two Micron All Sky Survey (2MASS) Catalog}\footnote{\url{http://www.ipac.caltech.edu/2mass/}} 
\citep[][]{2006AJ....131.1163S}. All images have been visually checked, eliminating those where the targets or the reference stars are close to the borders of the frame, and where obvious biases were present.

\subsection{Radio data}
Regular 15\,GHz observations of \pg were carried out as part of a high-cadence gamma-ray blazar monitoring programme using the Owens Valley Radio Observatory (OVRO) 40~m telescope \citep{2011ApJS..194...29R}.  \pg was observed with a nominal twice-per-week cadence.

The OVRO 40\,m uses off-axis dual-beam optics and a cryogenic receiver with a 3 GHz bandwidth centered at 15 GHz. The two sky beams are Dicke switched using the off-source beam as a reference, and the source is alternated between the two beams in an ON-ON fashion to remove atmospheric and ground contamination. In May 2014, a new dual-beam correlation receiver was installed on the 40~m telescope and the fast gain variations are corrected using a 180 degree phase switch instead of a Dicke switch. The performance of the new receiver is very similar to the old one and no discontinuity is seen in the lightcurves  (see Fig.~\ref{Fig:MWL_LC}). Flux density calibration is achieved using a temperature-stable diode noise source to remove receiver gain drifts and the flux density scale is derived from observations of 3C~286 assuming the \citet{1977A&A....61...99B} value of 3.44~Jy at 15.0~GHz. The systematic uncertainty of about 5\% in the flux density scale is not included in the error bars.  Complete details of the reduction and calibration procedure are found in \citet{2011ApJS..194...29R}.

Radio observations were also carried out with the 32-m dishes located in Medicina (at 8 and 24\,GHz) and  Noto (at 5\,GHz).  Continuum acquisitions were performed exploiting On-The-Fly cross-scans in Equatorial coordinates. 
Flux density calibration was carried out observing 3C 286, 3C 123, NGC 7027 and, prior to January 2018, also 3C 48. Reference flux densities for the calibrator sources were computed for the observed band central frequency, according to \citet{2013ApJS..204...19P}. 
For 24-GHz observations, the atmospheric contribution was also taken into account in the calibration procedure; the zenith opacity was estimated by means of skydip acquisitions.
The data reduction was performed using the Cross-Scan Analysis Pipeline described in \cite{2020MNRAS.492.2807G}.

\section{Characterization of the variability}
\begin{figure*}
\vspace{-2.5cm}
\centering
\includegraphics[width=1.\textwidth]{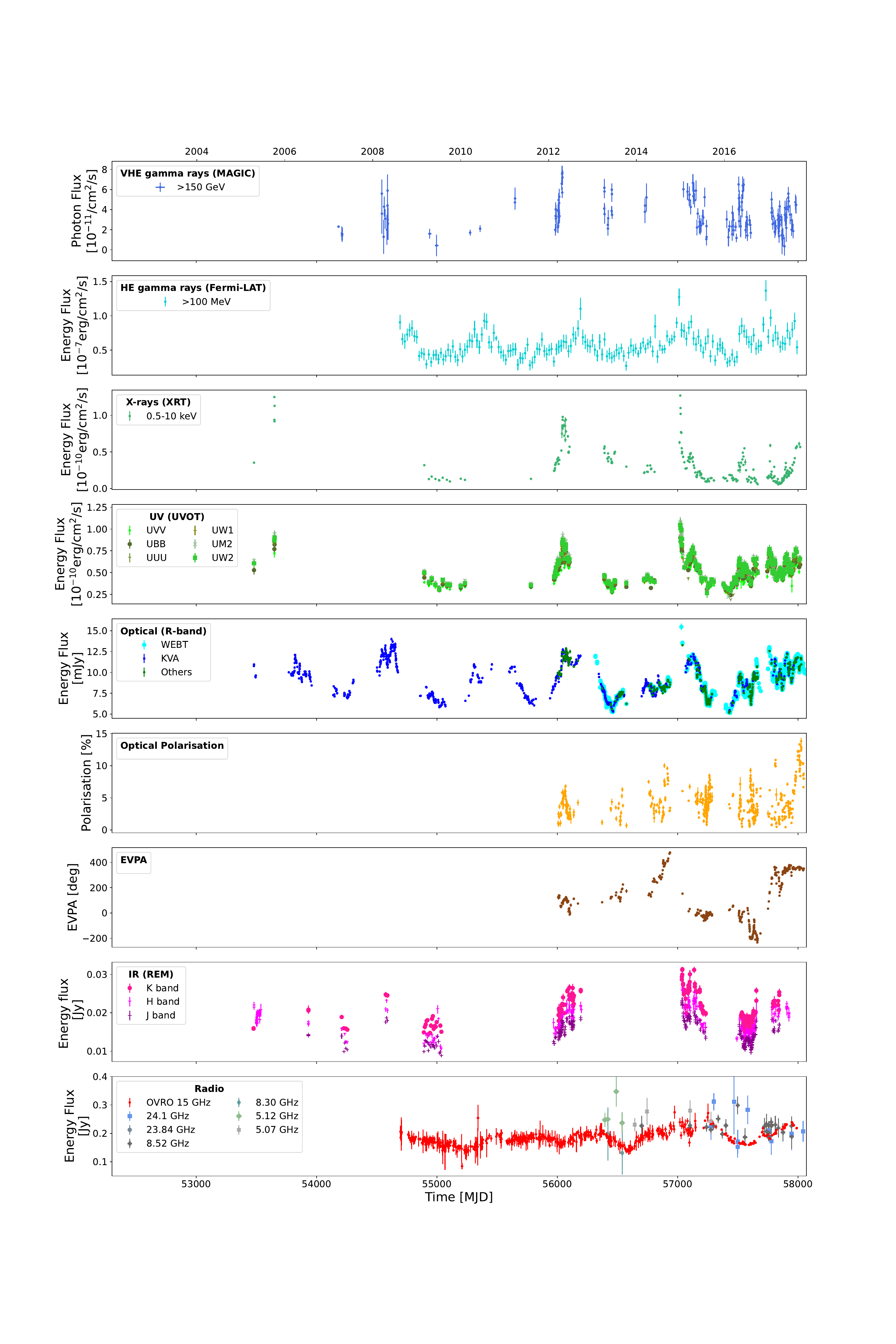}
\vspace{-3cm}
\caption{Long-term PG~1553+113 MWL lightcurves. From top to bottom: VHE gamma rays $>$150\,GeV (MAGIC, daily); HE gamma rays $>$100\,MeV(\textit{Fermi}-LAT, 20 days binning, from \citealt{2018ApJ...854...11T}); X-ray 0.5-10\,keV(\textit{Swift}-XRT); UV in six different filters (\textit{Swift}-UVOT, same snapshots than \textit{Swift}-XRT); optical in R band (WEBT, KVA, and from optical polarisation telescopes labelled as "others"); Optical polarisation and EVPA (NOT, RoboPol, Liverpool, Crimea, and  Steward telescopes); Infrared in three different filters (REM); and Radio observations in seven different frequencies (OVRO and Medicina radio telescopes).}
\label{Fig:MWL_LC}
\end{figure*}

Figure~\ref{Fig:MWL_LC} displays the lightcurves collected from PG~1553+113 at several wavelengths from radio (bottom panel) to VHE gamma-rays (top panel), for 12 years, from 2005 to 2017 as described in the previous section.

A large part of the HE gamma-ray data dataset, as well as radio and optical datasets, are published in \citet{2015ApJ...802...65A} and were used for the periodicity analysis.

The only instrument considered in this work that performed a continuous monitoring is \fermilat. Also radio data have very good coverage, followed by optical data that suffered only from a few months break per year related to the visibility of the source. The coverage is more scattered for IR, UV, X-rays, and VHE gamma-ray data that are strongly affected by sparse sampling and often the observations are driven by a high state alert trigger, and therefore may be biased towards high states. In these bands, the coverage had a clear improvement starting from late 2014 (MJD $\sim$57000). This is the result of an intense MWL and multi-year campaign aimed at a precise monitoring of the source state for the detailed modelling of the source emission and periodicity.

From a visual inspection of Figure \ref{Fig:MWL_LC} we can conclude that the source shows high variability over the years in all bands, with moderate variations in radio and more pronounced variations in the other bands. This behaviour is quite common in HBLs \citep{2021MNRAS.504.1427A}. 

A detailed characterisation of the variability is the key to investigate the physical phenomenology responsible for the broad-band emission as detailed in \citet{2019Galax...7...28R} and references therein. In the following subsections, several variability studies are presented. The aim is twofold: first, the characterisation of the variability (and periodicity) at different bands. Secondly,  the identification of inter-band  connections. These connections are a powerful tool to unveil single/multiple regions responsible for the observed emission.

\subsection{Flux-spectral index correlation}
A clear correlation between the integral flux and the slope of the power law approximating the differential energy flux in X-rays and gamma rays characterises many flaring events of BL Lacs. In the case of negative correlation, the effect is often referred as harder-when-brighter behaviour \citep[see, e.g.,][]{2007ApJ...669..862A}. This behaviour is associated to a shift of the synchrotron peak during flares, meaning that more energetic electrons are responsible for the bulk of the emission \citep[see e.g.][]{2021MNRAS.504.1427A}.

\begin{figure*}
\centering
\includegraphics[width=0.49\textwidth]{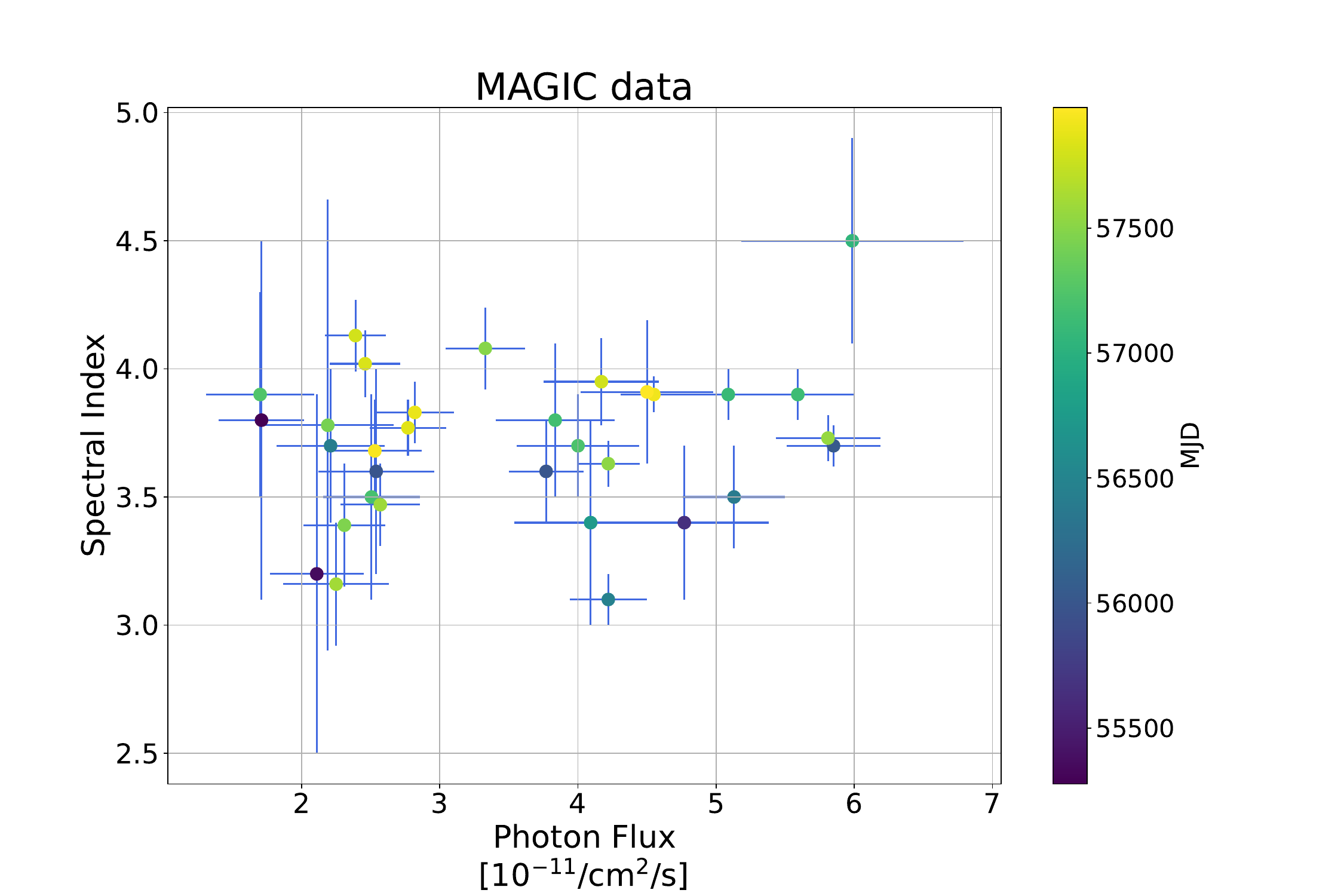}
\includegraphics[width=0.49\textwidth]{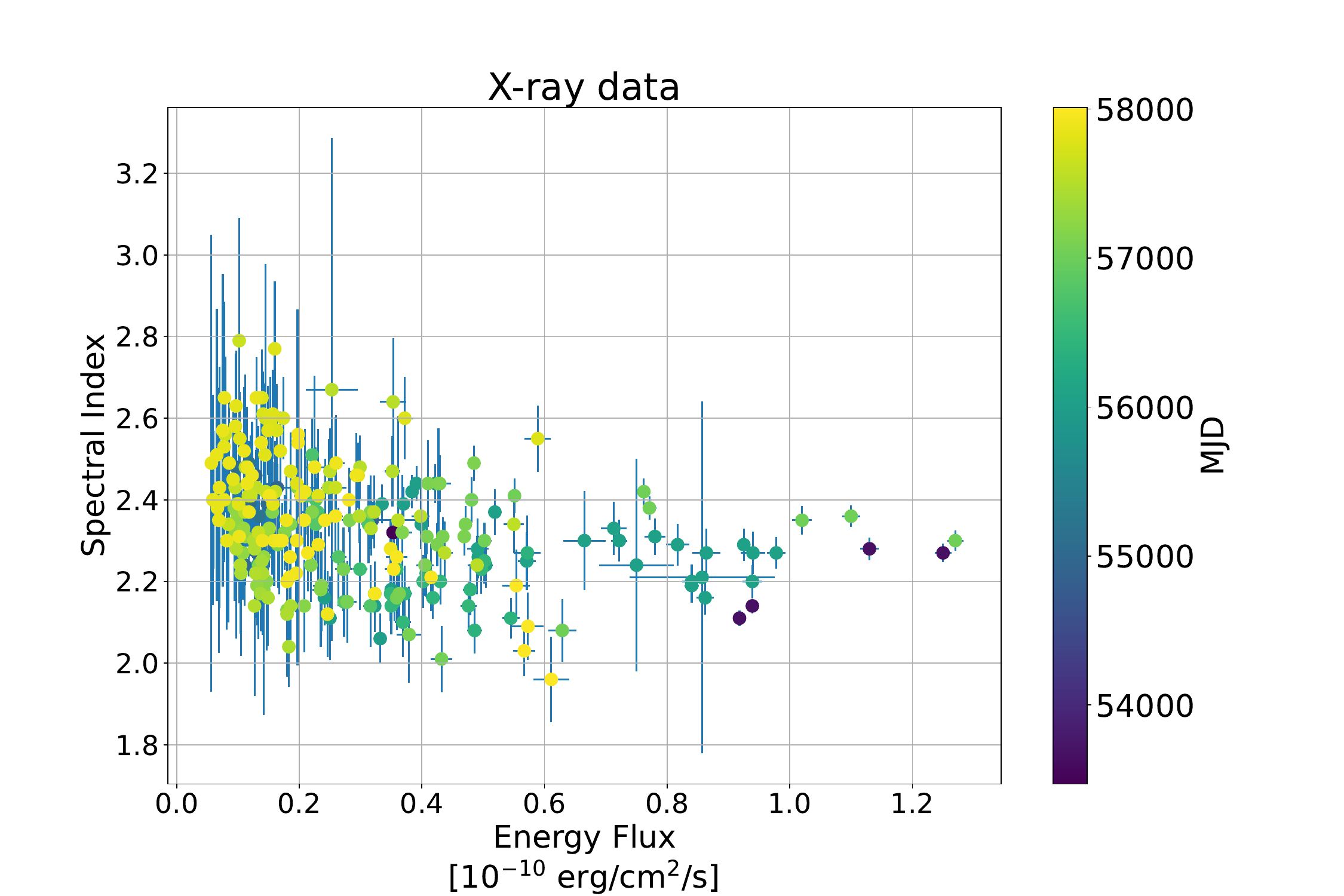}
\caption{Correlation study between the flux level and the spectral index of the power-law fit to the data above 150 GeV (left, MAGIC monthly averaged) and in the 0.5-10\,keV band (right, \textit{Swift}-XRT individual pointing).}
\label{fig:flux-slope}
\end{figure*}

A study of the flux-slope correlation was performed for the VHE gamma-ray and X-ray data. The results are shown in Fig.~\ref{fig:flux-slope}. The monthly averaged MAGIC data from 2010 to 2017 were considered for the study. No correlation appears in the MAGIC data, but the large time interval considered and the relatively low statistics involved in the study may have diluted this correlation. X-ray data show instead a hint of anticorrelation between the spectral index and the flux state, indicating a harder-when-brighter trend.
To evaluate the level of correlation we adopted the Spearman correlation coefficient, a value close to (-)1 pointing to a strong (anti-)correlation. The Spearman correlation coefficient in X-ray data is -0.39, and the p-value of the null-hypothesis (i.e., no-correlation) is $\sim 10^{-10}$.

It is important to emphasise that, contrary to the vast majority of studies available in literature claiming a harder-when-brighter trend in the considered bands \citep[e.g.,][]{1998ApJ...492L..17P,2007ApJ...669..862A,2020ApJS..247...16A}, the results presented here are not from a single campaign/flare, but they come from a large time interval spanning more than 10 years. 

\subsection{Variability timescale}
Apart from few exceptions \citep[e.g.,][]{2020ApJS..247...16A}, blazars are highly variable objects in almost all bands. In HBLs, the variability timescale may range from months down to a minute timescale. 
According to special relativity, the comoving size of the emitting region, $\Delta r'$, can be constrained by the variability timescale, \citep[as detailed in][]{2019Galax...7...28R}.
The variability timescale pinpoints the properties of the region responsible for the observed radiation.
Interestingly, subday variations represent a challenge for the simplest emission models in blazars \citep[e.g.][]{1998ApJ...509..608T}, and for the subclass of Flat Spectrum Radio Quasars \citep[e.g., fast variability--10 minutes doubling time--observed in VHE gamma rays in PKS~1222+21, ][]{2011ApJ...730L...8A}.

To constrain the variability timescale of \pgn, a search for intraday variations was performed on MAGIC and \textit{Swift}-XRT data. For statistical reasons, the study was limited to 10 snapshots with the highest flux recorded in both bands. 
The average duration of MAGIC observations was 1\,hour, while the average duration of \textit{Swift}-XRT observations was 1.2\,ks. The analyses revealed no hint of intraday variability in \textit{Swift}-XRT and MAGIC data. 

\subsection{Interband correlation}
Short or long-term correlation between the fluxes emitted at different bands allows us to track down the 
connection between photons emitted at different possible regions in the jet or with different, but correlated, mechanisms. 
This is the case of synchrotron self-Compton emission (SSC), where low-energy synchrotron photons are emitted together with inverse Compton, high-energy photons produced by the same electrons upscattering the synchrotron photons.

In our study, we focus on the interband correlation search on the IR, optical, UV, X-ray, HE and VHE gamma-ray bands. Radio data are excluded from this study in consideration of the well known lag due to the different location of the emission zone, that will be further discussed in the text.

The results of the correlation analysis performed with the Spearman test as implemented in the \texttt{SciPy}  python package are listed in Tab.~\ref{Tab:Correlation}, where the Spearman coefficient and the p-value appear in the third and fourth columns, respectively.

\begin{figure*}
\centering
\includegraphics[width=0.46\textwidth]{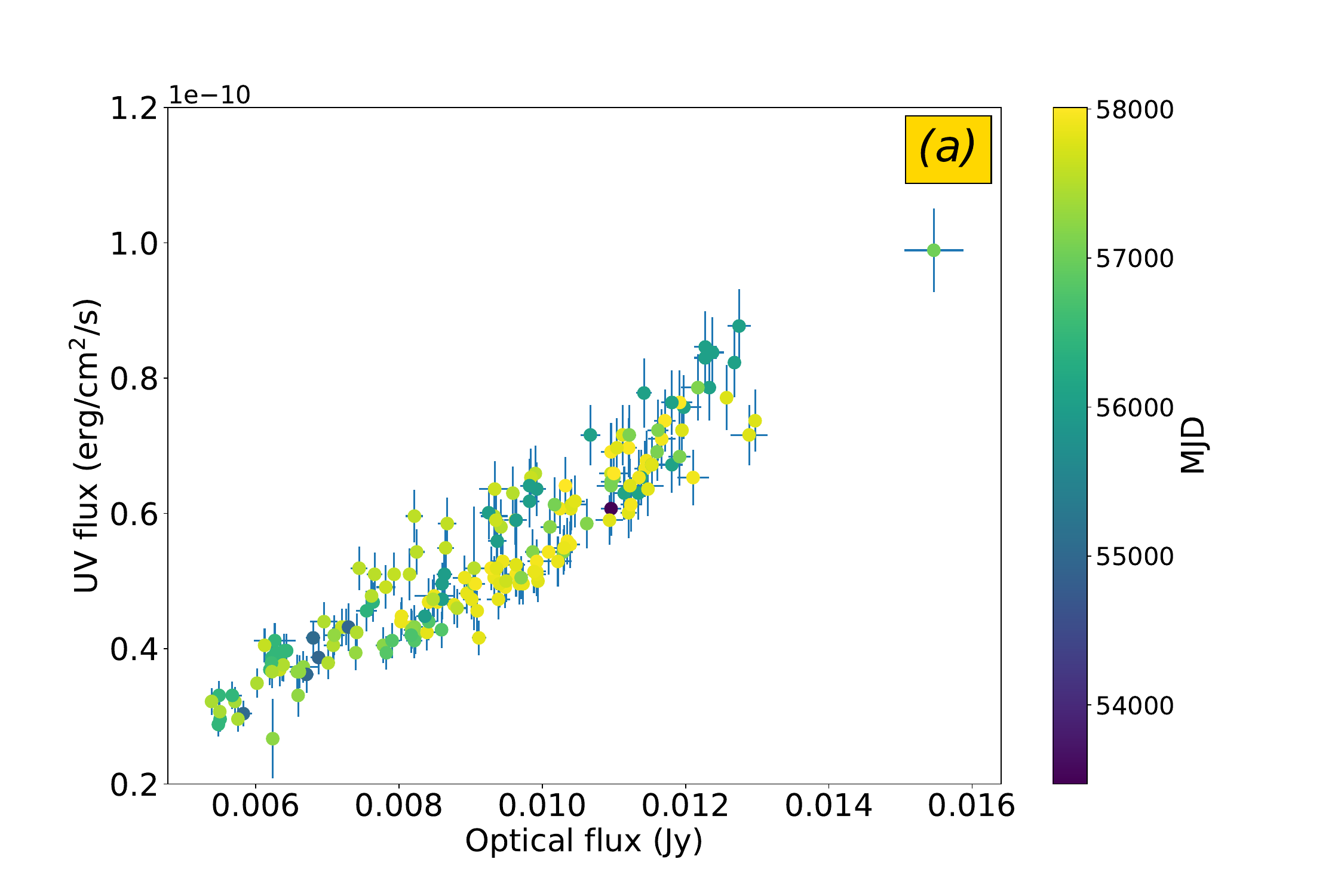}
\includegraphics[width=0.44\textwidth]{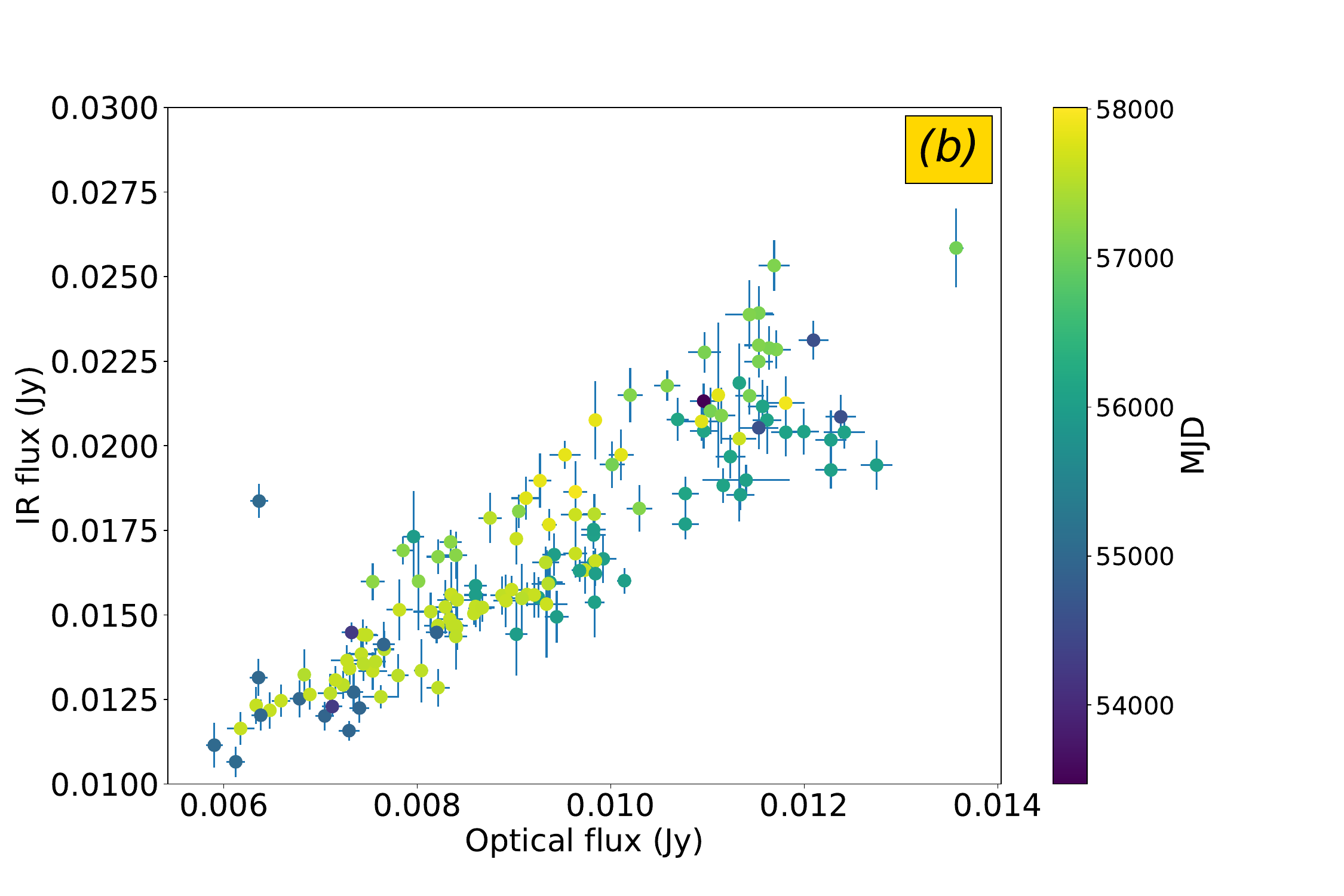}
\includegraphics[width=0.46\textwidth]{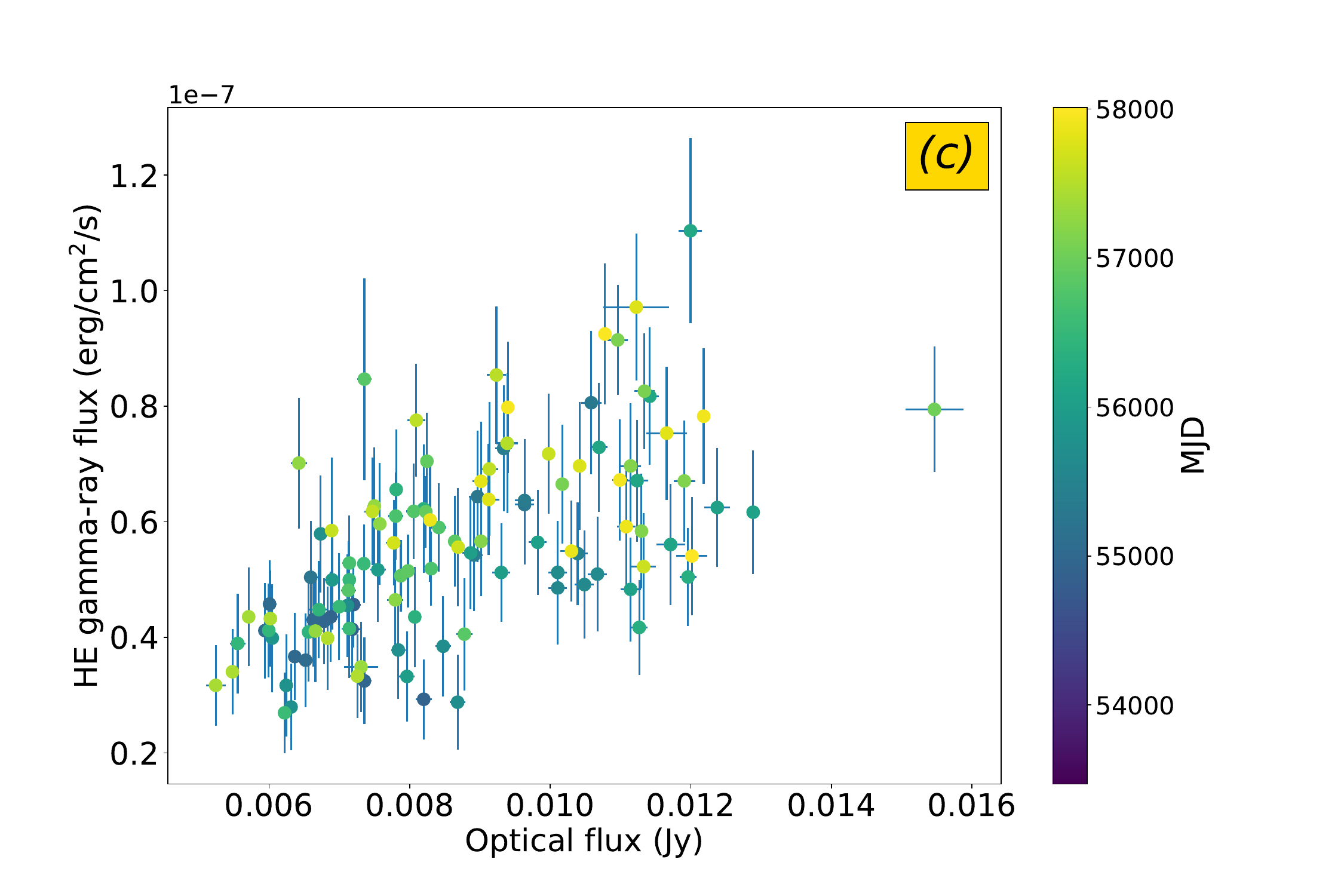}
\includegraphics[width=0.46\textwidth]{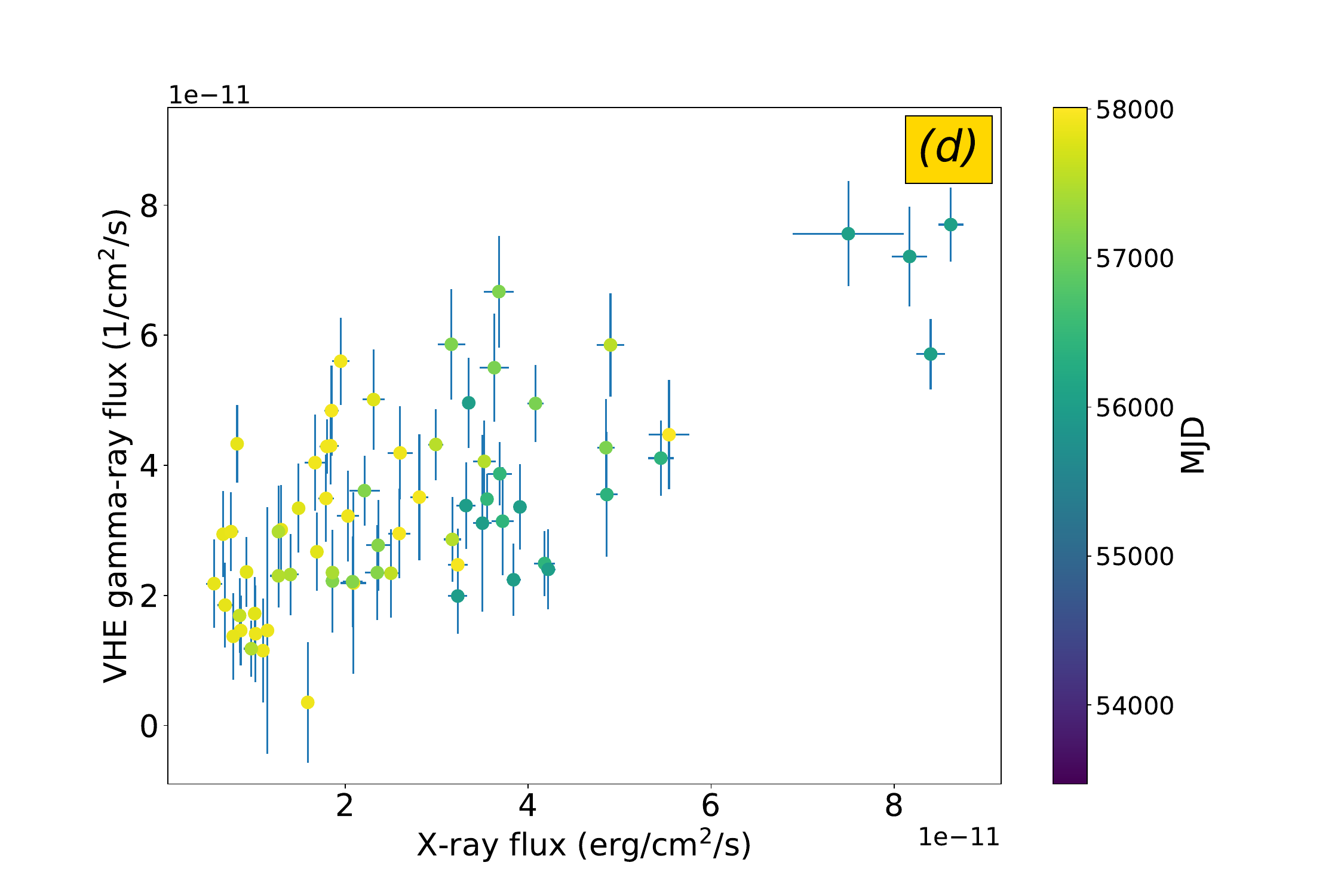}
\includegraphics[width=0.46\textwidth]{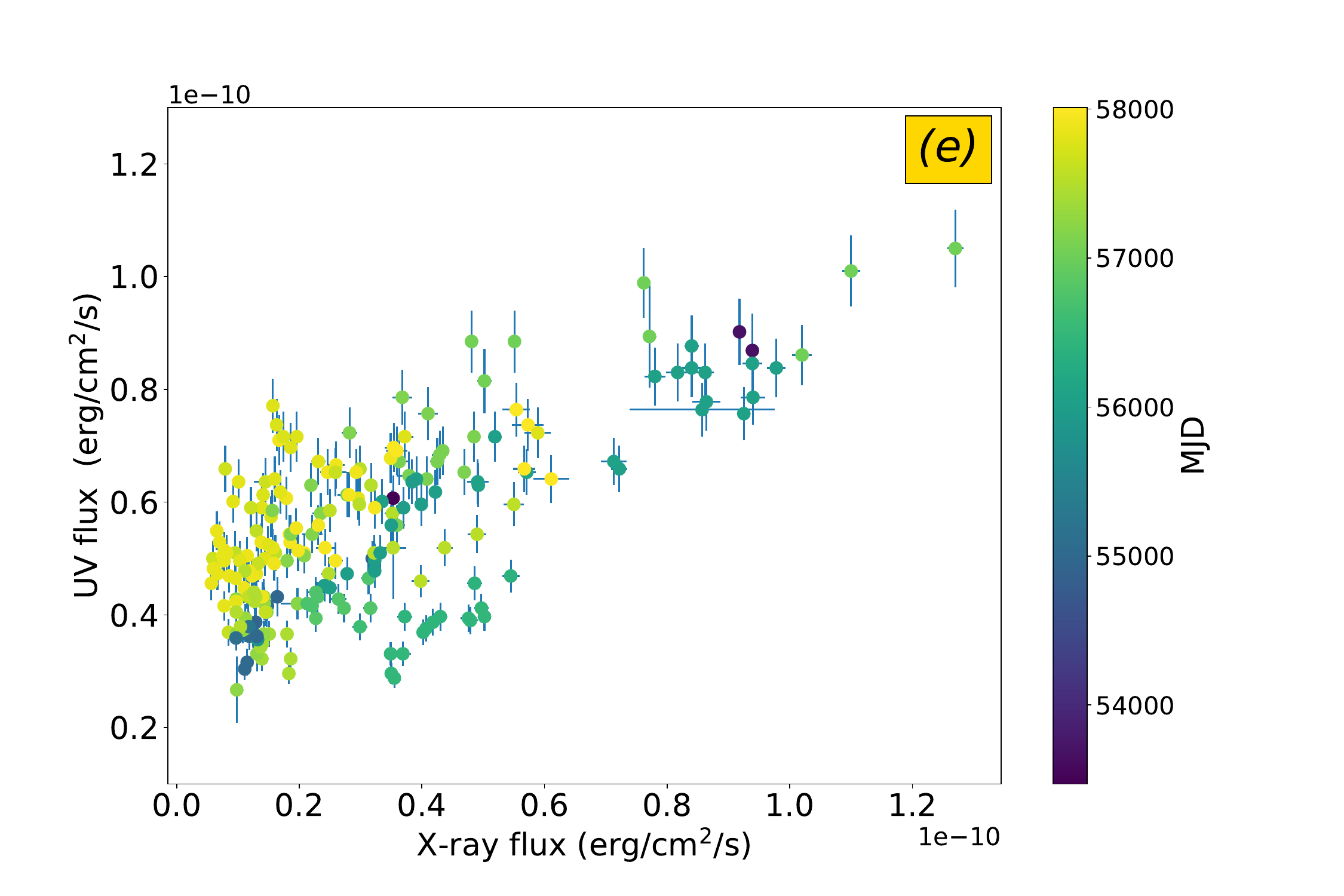}
\includegraphics[width=0.46\textwidth]{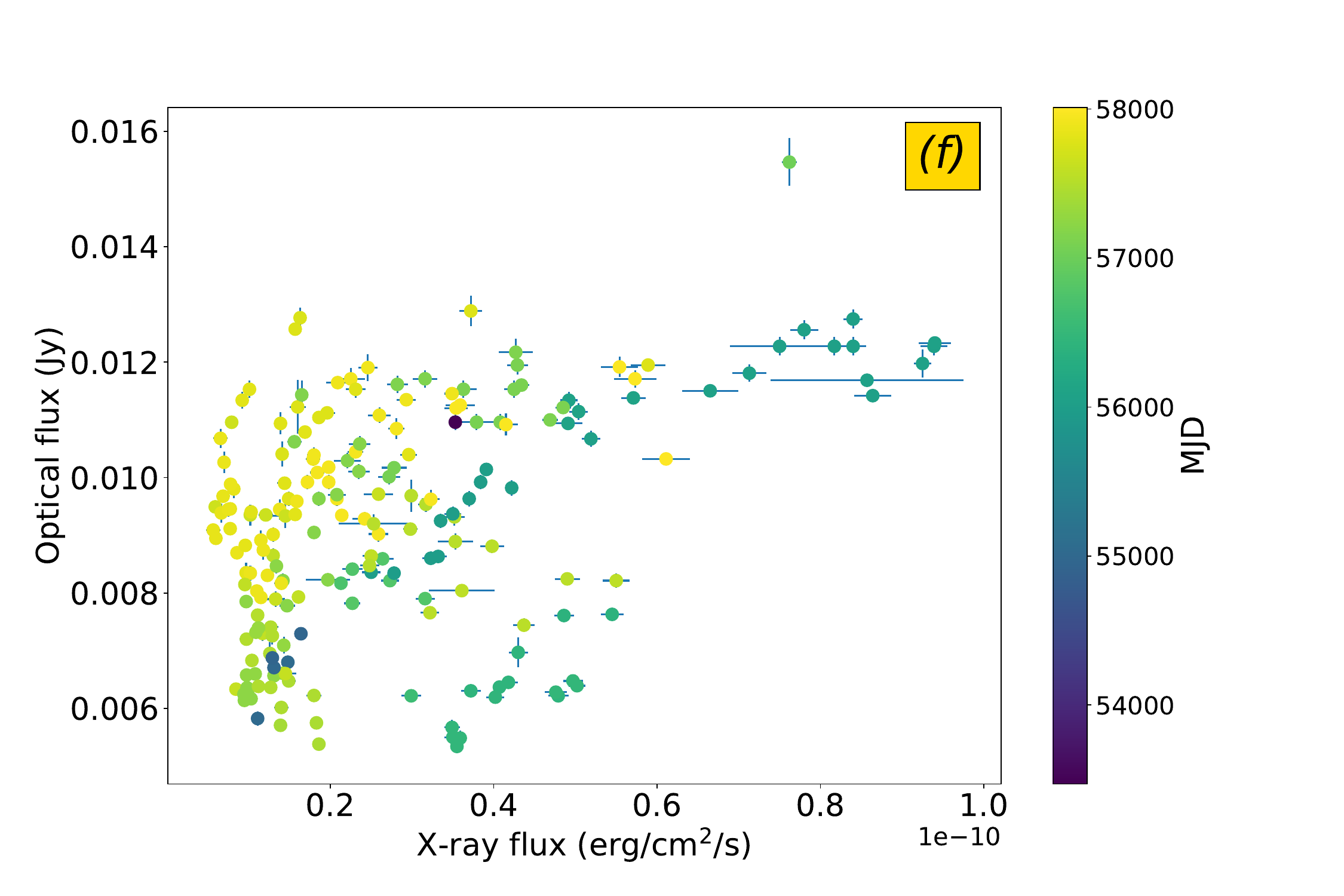}
\caption{Selected scatter plots used to investigate flux correlations. For correlation studies involving \textit{Fermi}-LAT, the simultaneity window has been set to $\pm$ 10 days. In all other cases, the window assumed for simultaneity is $\pm$1.5\,days.}
\label{fig:correlation}
\end{figure*}

\begin{figure}
\centering
\includegraphics[width=0.46\textwidth]{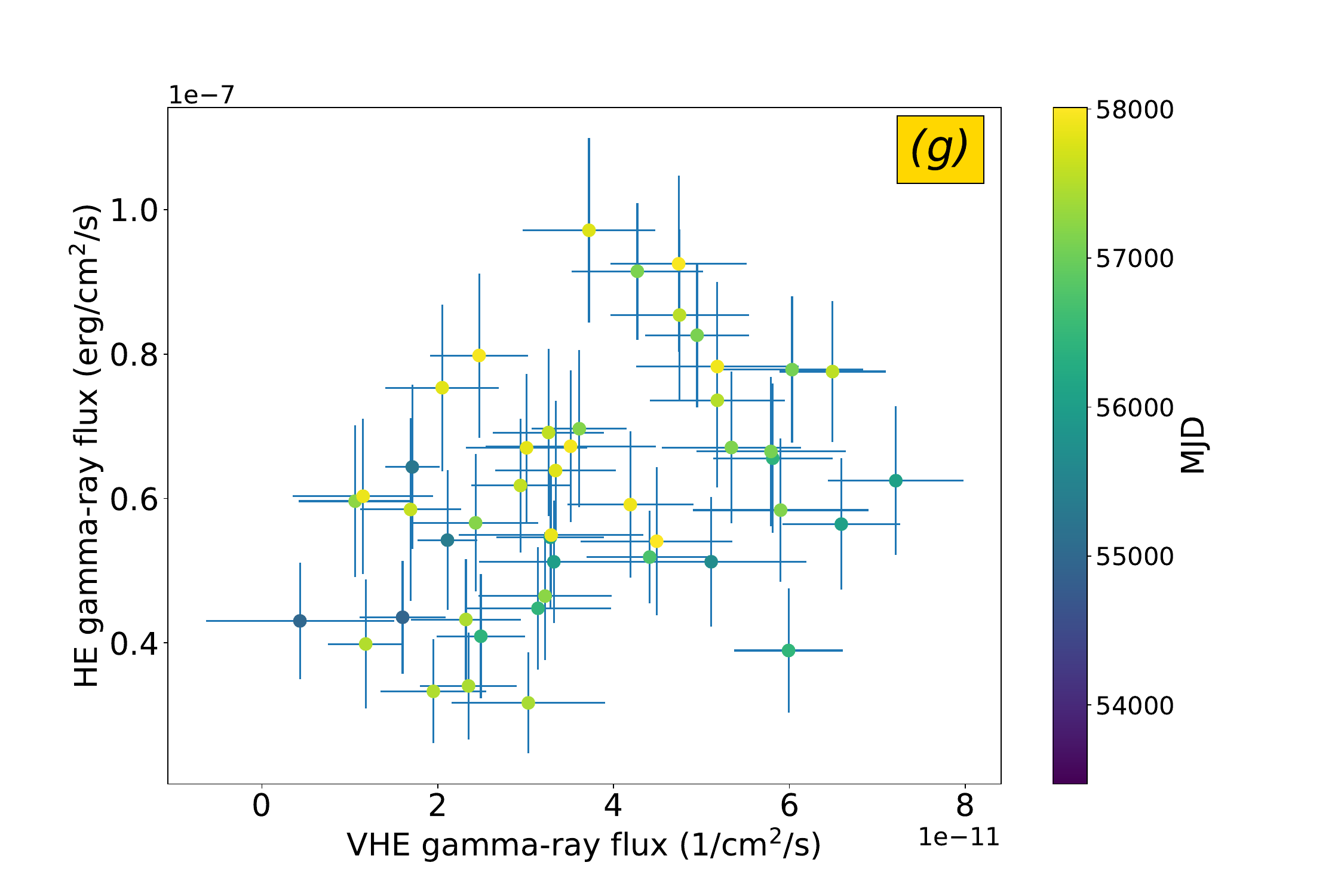}
\includegraphics[width=0.46\textwidth]{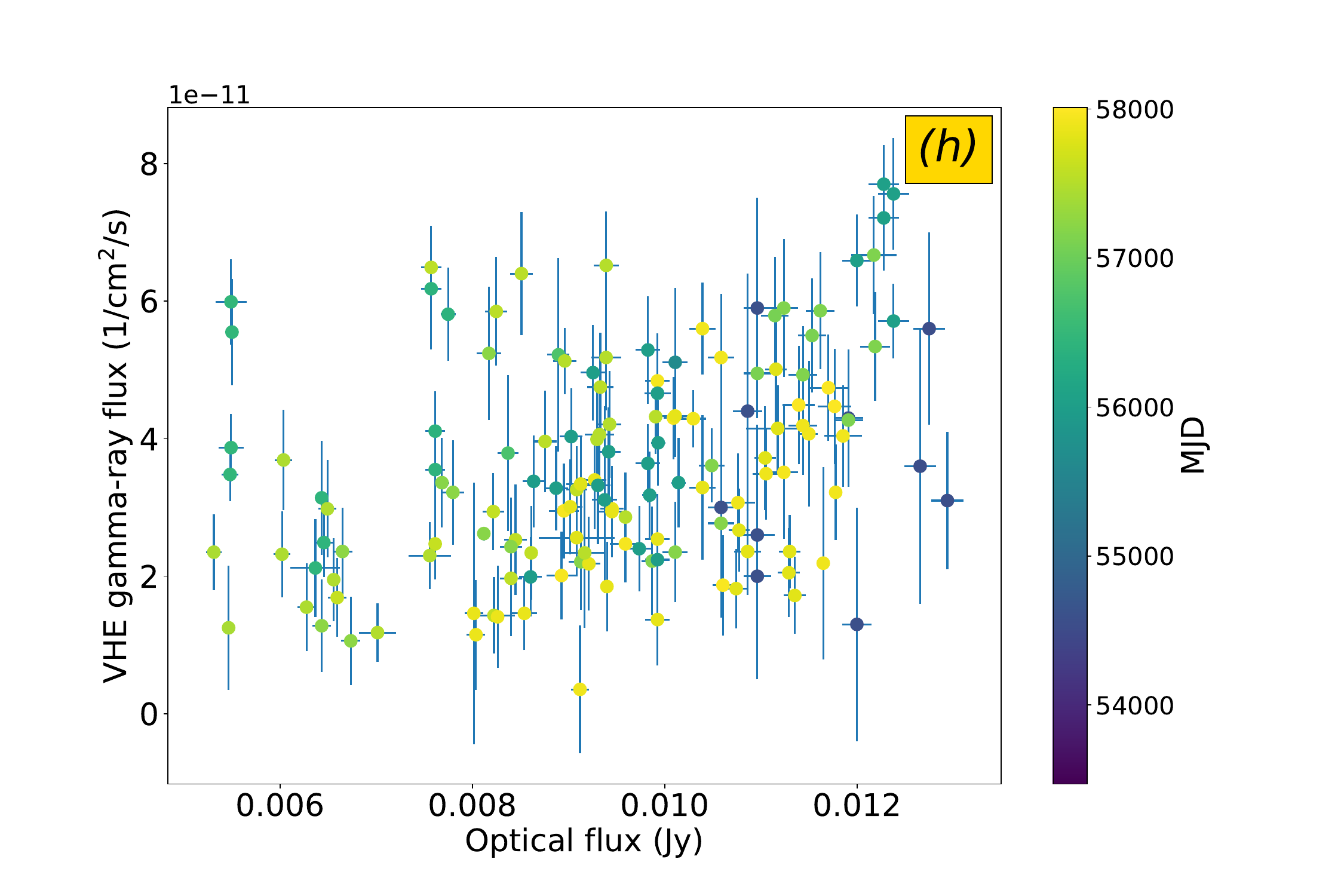}
\includegraphics[width=0.46\textwidth]{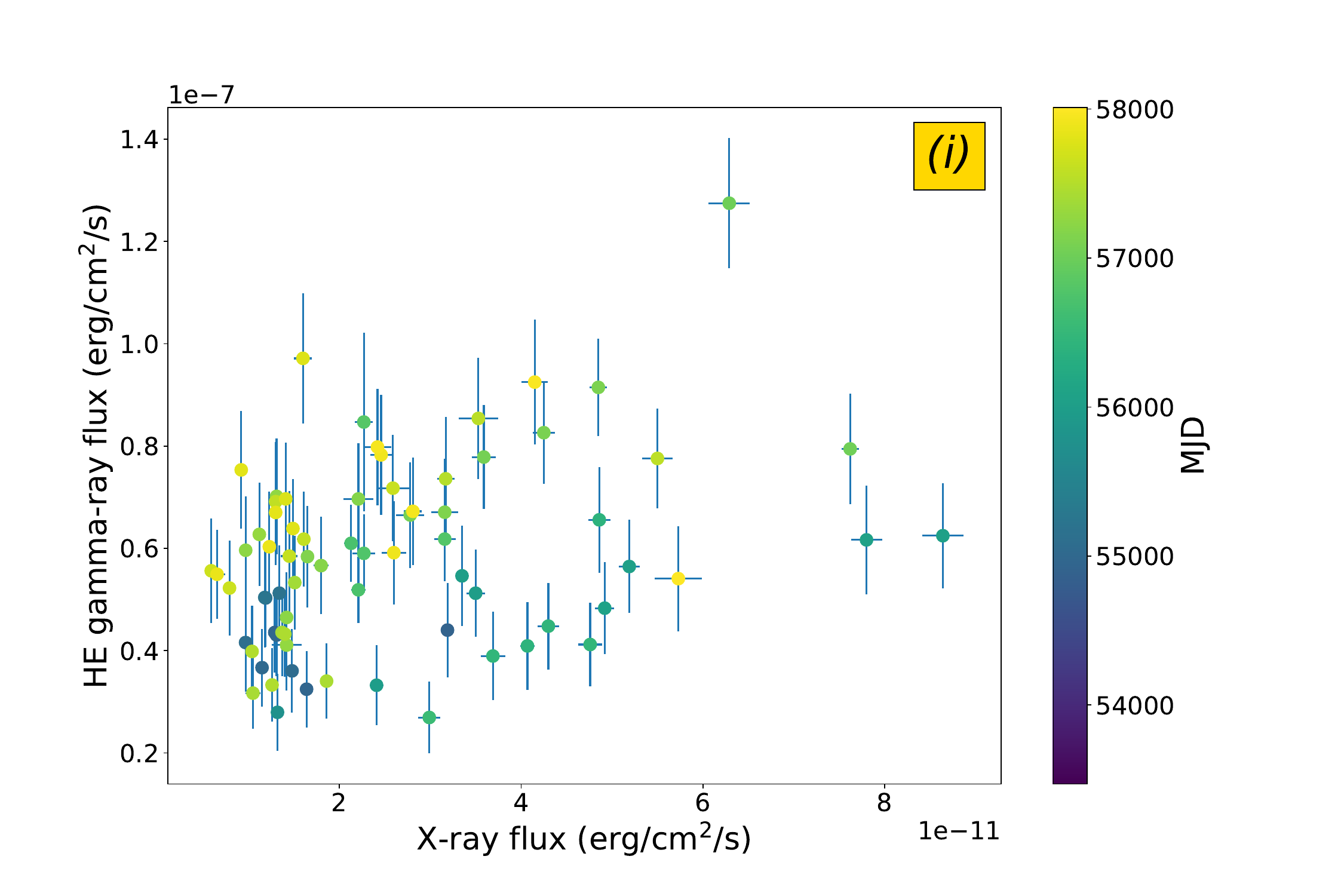}
\caption{Continue from Fig.~\ref{fig:correlation}.}
\label{fig:correlation2}
\end{figure}

For this study, only data within a 1.5-day window have been considered simultaneous apart for the correlation studies involving \fermilat\, data where the simultaneity window has been extended to $\pm$~10 days, to have sufficient statistics and in agreement with the \fermilat\, data binning of 20 days.
The Table is ordered by decreasing the correlation levels according to this indicator. Similar results are obtained with the weighted Pearsons coefficient (that has the advantage of taking into account the errors of the flux, but the disadvantage of assuming a Gaussian distribution of values).
 Fig.\,\ref{fig:correlation} and \ref{fig:correlation2} show some selected scatter plots, also indicated in the last column of Table~\ref{Tab:Correlation}.

\begin{figure}
    \centering    
    \includegraphics[width=0.52\textwidth]{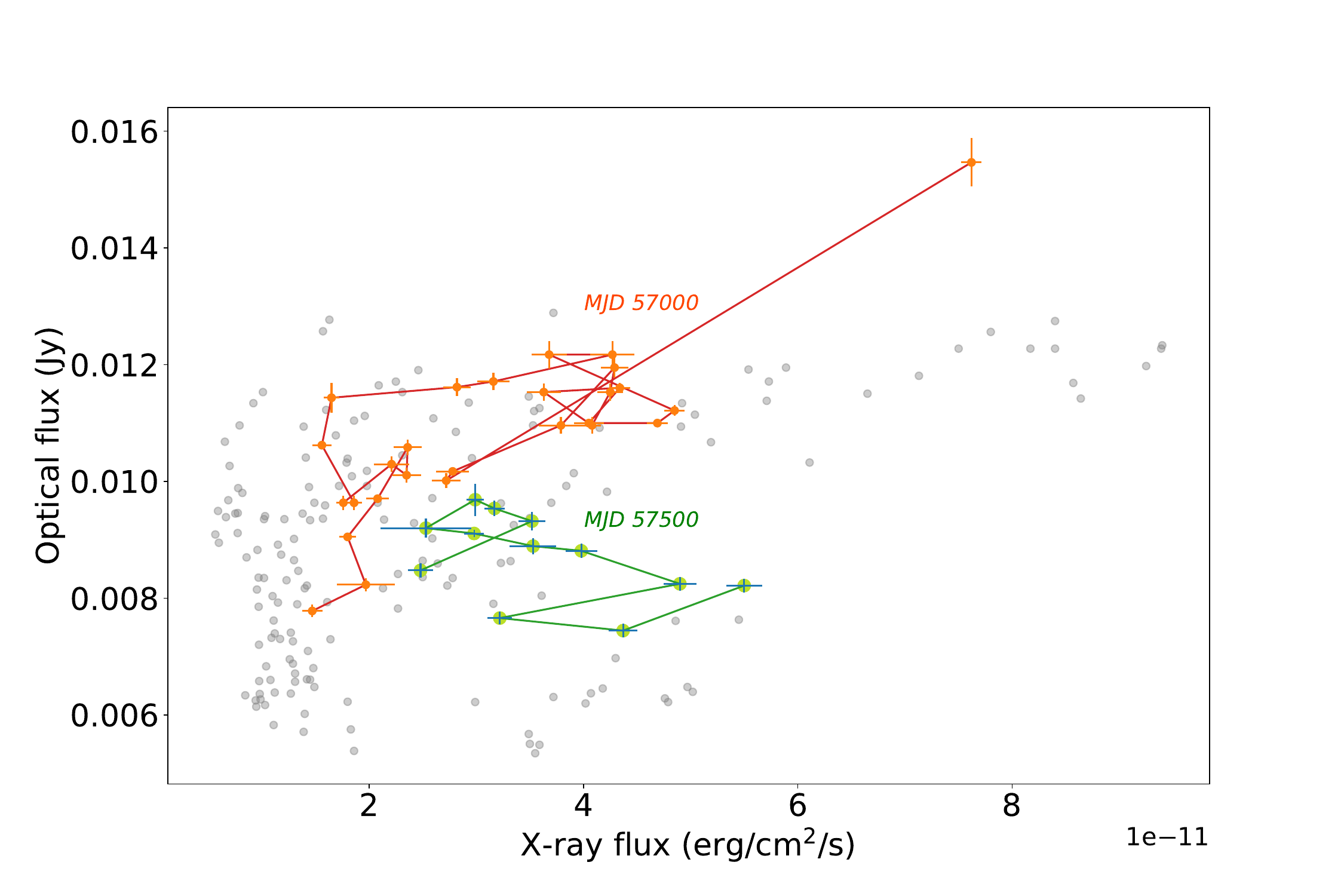}
    \vspace{0.2cm}
    \caption{Two selected episodes with different correlation behaviour, correlation (MJD~57000) and anti-correlation (MJD~57500), in the optical-X-ray scatter plot.}
    \label{fig:corelation_opt-xray_selected}
\end{figure}

The main results of the correlation analysis are as follows:
\begin{itemize}
    \item[-] A strong correlation (Spearman coefficient $\ge$\,0.9) is observed both between optical and UV (UW2 band) and optical and IR (H band) data;
    \item[-] Optical and UV data show a net correlation (Spearman coefficient $\ge$\,0.6) also with the HE gamma ray data. A similar relation is also observed between X-rays and VHE gamma-rays;
    \item[-] X-ray and UV data, with strictly simultaneous sampling, show a milder correlation (Spearman coefficient $=$\,0.55);
    \item[-] Only a hint of correlation (Spearman coefficient $\le$\,0.4) with close to zero time lag is observed between the other bands (HE and VHE gamma rays, optical/X-ray, and optical VHE gamma rays). 
    \item[-] In the optical/X-ray case, a careful analysis of the scatter plot allows us to identify episodes with different correlation behaviours: from a clear correlation, corresponding to the X-ray and optical flare at MJD $\sim$ 57000 to anti-correlated events (X-ray enhanced state at MJD $\sim$ 57500), as highlighted in Fig.~\ref{fig:corelation_opt-xray_selected}. \end{itemize}

These results suggest a common origin for the spectral features observed in IR, optical, UV,  and \fermilat\, bands. In particular, IR, optical, and UV photons are likely synchrotron photons from the same emitting region. Single-zone SSC process  is the cogent mechanism connecting optical and HE gamma-ray photons.  
The same process may be responsible for the X-ray to VHE gamma-ray connection, even if the radiation should come at least in part from a different (or additional) region with respect to the low-energy counterpart, to explain the weaker correlation with the other bands. 

The possibility of a delay in the \pg correlation between bands has recently been investigated in \citet{2018MNRAS.480.5517L} for the radio, optical, and gamma-ray bands. Although gamma-ray and optical data are consistent with no time-lag correlation, a delay of $\sim$3-4 months appears between radio and both optical and \fermilat\ data. 
We have investigated the possibility of a delay between radio and X-ray and VHE gamma-ray data, between  \fermilat\ and X-ray and VHE gamma-ray data, and between X-rays and VHE gamma rays with the same method presented in \citet{2018MNRAS.480.5517L}. In all cases, from the discrete correlation function study no significant time lag emerged.
The analysis of radio and optical/gamma-ray data, instead, are fully consistent with those reported in \citet{2018MNRAS.480.5517L}. A delayed correlation in the radio band is well known effect in blazars and is due to the self-absorption of radio emission, which becomes visible when the density of the region drops, inducing a delay with respect to high-energy emission.

\begin{table}
\centering
\caption{Results of the correlation study between integral flux in different bands ordered by decreasing Spearman coefficient (third column).  The simultaneity window assumed is $\pm$1.5-day apart for the correlation studies involving \fermilat\, data,  where it has been extended to $\pm$ 10 days. The last two columns report the p-value (null hypothesis: no correlation with close to zero time lag) and, if available, the panel with the scatter plot in Fig.\ref{fig:correlation}, respectively.}
\begin{tabular}{l|l|c|c|c}
\hline
\hline
Band-1 & Band-2  & Spearman & p-value & panel\\ 
&  & Coeff. & & \\
\hline
\hline
Optical & UV   & 0.94 & 4e-88  & a \\ 
Optical & IR  & 0.90 & 2e-50 & b\\ 
UV & HE $\gamma$-ray    &  0.66 & 3e-10 & \\ 
Optical & HE $\gamma$-ray & 0.63 & 2e-14 & c\\ 
UV & VHE $\gamma$-ray &  0.62 & 9e-08 & \\ 
IR & HE $\gamma$-ray & 0.61 & 1e-05  & \\ 
X-ray  & VHE $\gamma$-ray   & 0.60 & 6e-08 & d\\ 
IR & UV & 0.60 & 4e-06 & \\ 
UV & X-ray &  0.55 & 6e-18 & e\\ 
Optical  & X-ray    & 0.37 & 4e-08 & f\\ 
HE $\gamma$-ray  & VHE $\gamma$-ray  & 0.39 & 0.006 & g \\ 
Optical & VHE $\gamma$-ray & 0.35 & 2e-05 & h\\ 
X-ray & HE $\gamma$-ray   & 0.32 & 0.006 & i\\ 
IR & VHE $\gamma$-ray & 0.26 & 0.09 & \\  
IR & X-ray & 0.29 & 0.02 & \\ 
\hline
\hline
\label{Tab:Correlation}
\end{tabular}

\end{table}

\subsection{Bayesian block analysis}
\begin{figure*}
\centering
\includegraphics[width=0.9\textwidth]{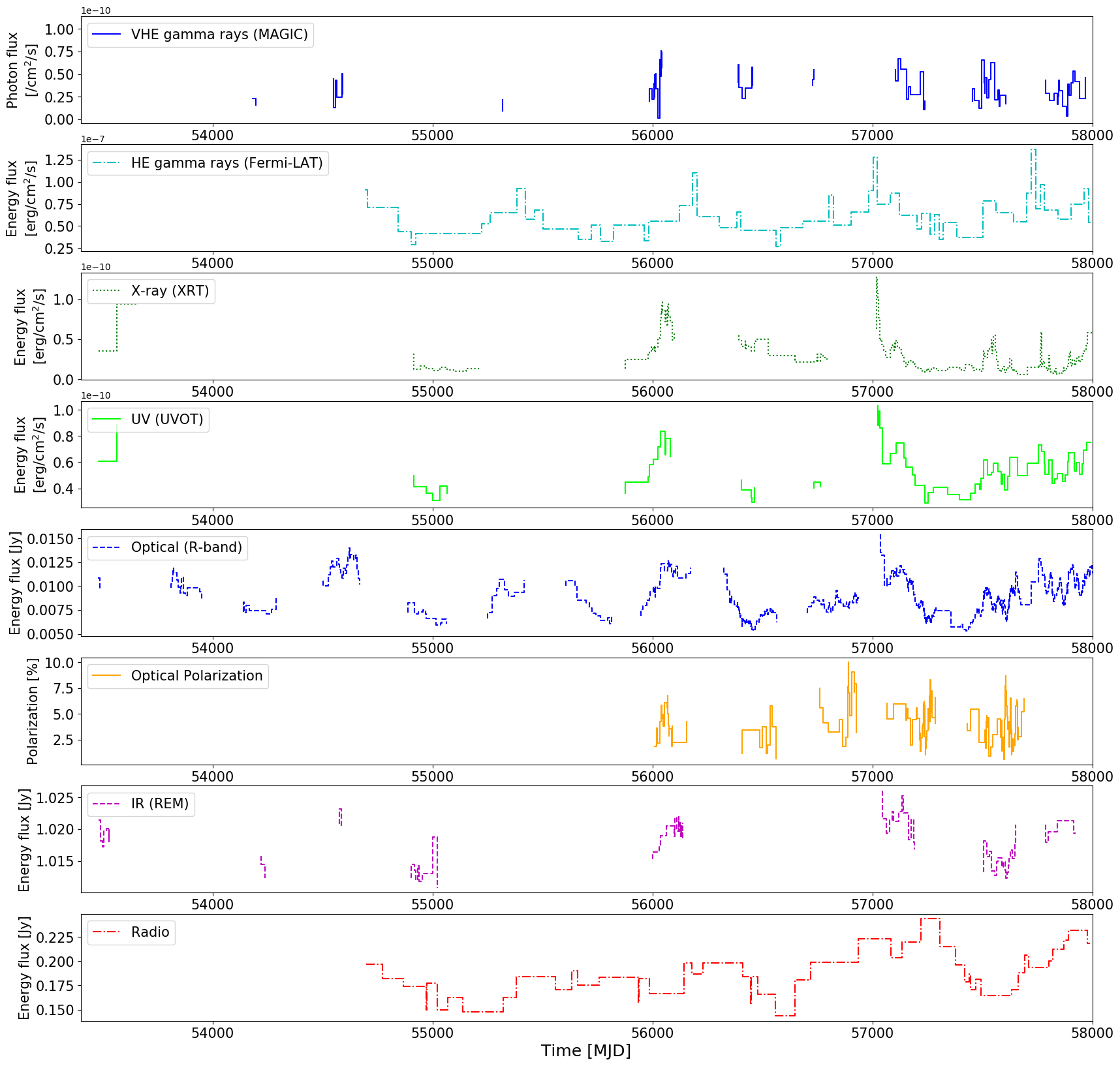}
\caption{Bayesian block representation of the MWL lightcurves having excluded the observational gaps.}
\label{fig:bayesian_blocks}
\end{figure*}

In order to determine variations and flares in the MWL light curves, we model flux variations in a model-independent manner using Bayesian blocks \citep{2013ApJ...764..167S}. Figure \ref{fig:bayesian_blocks} shows the  Bayesian block representation of the MWL lightcurves, including the optical polarisation degree. We have optimised the prior of the slope on the number of bins for individual lightcurves to better match their sampling and variability.

We find interesting flaring behaviour across all wavelengths. For example, we find contemporaneous flares in X-rays, UV,  optical, and elevated activity in VHE gamma rays and polarisation at $\sim$56060~MJD that does not seem to have a counterpart in the \fermilat\ band. On the other hand, we find contemporaneous flares in optical, UV, X-rays and gamma-rays at $\sim$57020~MJD. The 15~GHz radio also seems to be in an elevated state during that time, although the radio lightcurve does not follow the variability patterns in other wavelengths. Unfortunately, apart from $\sim$57020~MJD, the prominent {\it Fermi} flares fall into optical and/or VHE gamma-ray gaps, although in all of them the optical seems to be in an elevated state.

\section{Periodic modulation in the MWL lightcurve}\label{sect:periodicity}

As first proposed in \citet{2015ApJ...813L..41A} and then confirmed in several other studies \citep{2020ApJ...895..122C,2018A&A...615A.118S,2017MNRAS.471.3036P,2016MNRAS.458.1127G}, a periodic modulation of the HE gamma-ray emission is firmly established in \fermilat\ data. 
Although the optical curve appears to be much more complex than \fermilat\ curve, the modulation with a period similar to that observed at higher frequencies is found, with a smaller significance  \citep{2018ApJ...854...11T}.

\begin{figure}
\vspace{-3cm}
\centering
\includegraphics[width=0.45\textwidth]{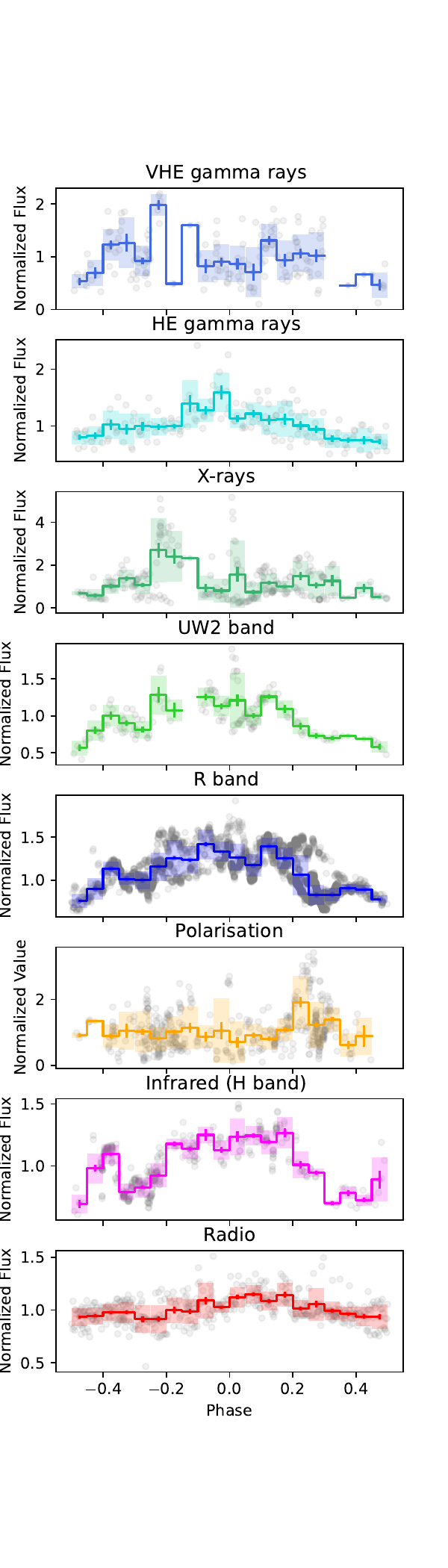}
\vspace{-3cm}
\caption{PG~1553+113 MWL folded lightcurves obtained by assuming a period of 798 days. In each panel, the average values are reported with a continuous line. The error bar in each bin is the standard error of the mean values. The colored bar in each bin represents the variance of the data distribution.} 
\label{Fig:Folded_LC}
\end{figure}

As a first step in the periodicity study, we visually inspected if the lightcurves at different bands are in agreement with the hypothesis of a periodic modulation of period P$_{fermi}$ = 798\,days.
To this purpose, we built a normalised MWL folded lightcurve assuming a period of 798 days \citep{2015ApJ...813L..41A}. The final result is shown in Fig.~\ref{Fig:Folded_LC}  obtained for all data reported in Fig~\ref{Fig:MWL_LC} for the radio, optical polarisation, optical, UV, X-ray, HE gamma-ray, and VHE gamma-ray bands. 
In each bin, the average value and its error are reported with a continuous line. The bars instead represent the square root of the variance of the folded data of each phase bin.
We underline that, while the error on the average is strongly dependent on the number of points in that bin, the variance is not. Therefore, the latter is an indication of the dispersion of the sample, except for a few cases with bins with a single point (hence with a low uncertainty that is by no means representative of the real dispersion).

As expected, the HE gamma-ray folded curve displays a clear peak at phase $\sim$\,0. The minimum emission is instead located at phase 0.3--0.5. The same trend is observed in the optical and UV curves, while the 
radio curve behaviour seems shifted, 
in agreement with the delayed emission resulting from the DCF analysis in \citet{2018MNRAS.480.5517L}, also confirmed in our study. 

It is interesting to note that, despite the poor sampling of the lightcurves, X-ray, and VHE gamma ray-folded curves are almost fully characterised, meaning that the observations carried out in the 11 years considered were diluted in different phases of the assumed periodic modulation. This allows us a first qualitative study of the trend of these curves. The main finding of this study is that while the folded curves, including the optical polarisation, show a minimum in the phase interval 0.3-0.5, in agreement with the observations in the HE gamma-ray band, no clear maximum is observed in the optical polarisation, X-ray, and VHE gamma-ray bands.  

In parallel with this visual inspection, we have also carried out a study of periodicity, which is reported in the next section.

\subsection{Systematic search for periodicity}

In all bands we search for a sinusoidal periodic signal using the \ac{gls} periodogram \citep{2009A&A...496..577Z} as implemented in the \verb|PyAstronomy| python package\footnote{\url{https://github.com/sczesla/PyAstronomy}} \citep{2019ascl.soft06010C} with frequencies ranging from $1/T$ to $N/(2T)$ sampled in steps of $1/(10T)$, where $T$ is the total time of the light curve and $N$ is the number of data points.

We identify the period and power of the strongest peak. In the periodogram of the radio data the \ac{gls} power at the lowest frequency is higher than the next strongest peak; however, we do not take the low-frequency peak into consideration because the peak period lies outside of the covered frequency range and less than one full cycle would be covered by the full data.

We need to assess whether a detected peak provides significant evidence for an intrinsic periodicity or whether it is a sporadic result of the typical flaring behaviour. We follow the procedure described in Appendix~A of \citet{2022ApJ...926L..35O}.
Our Null hypothesis is that the light curves follow a stochastic red-noise process, with the same statistical properties of the original data -- namely the \ac{psd} and \ac{pdf}.
We assume a power-law \ac{psd} $\sim \nu^{-\beta}$, where $\nu$ is the frequency, and estimate the index, $\beta$, with an implementation of the method introduced by \citet{2002MNRAS.332..231U}. In the radio band the index was found to be $2.0\pm0.51$ and for the MAGIC data $1.0\pm0.42$, for the remaining bands ranging from 1.4--1.5 with uncertainties ranging from 0.1--0.26.

We estimated the \acp{pdf} through the \acp{ecdf} of the light curves. For the {\it Fermi}-LAT analysis we created 50\,000 simulations, for all other bands 10\,000~simulations each, that implement the Null hypothesis.
We calculate the \ac{gls} for all simulations and count the simulations that have a power equal to or higher than the peak power at the peak frequency of the data \ac{gls}; this gives us the local p-value, i.e. the probability that a red-noise process results in an apparent periodicity as strong or stronger than observed at specifically the frequency where it was detected. The local p-value does not take into account that we test many frequencies. To take the look-elsewhere-effect into account we identify the strongest peak (discarding peaks at the edge of the frequency range) in the \ac{gls} of each simulation and calculate its local p-value, then we count all simulations with peaks that have a local p-value equal to or lower than the local p-value of the data \ac{gls} peak to estimate the global p-value.

Our results are reported in Table~\ref{tab:periodicity_results1}, where the PSD index, the peak period in days, the peak power and the local and global p-values are listed as a function of the considered band (first column).
The \ac{gls} of the \textit{Fermi}-LAT light curve shows a prominent peak at a period of 786~days, with a global p-value of $1.0\times10^{-3}$, which corresponds to $3.1\sigma$ in a one-sided test if the statistic were normal distributed. At an a~priori chosen significance level of $3\sigma$ we reject the Null hypothesis that the detected peak in the periodogram is a likely result of a red-noise process and conclude that light curve very likely contains a truly periodic signal. 

The \textit{Swift}-XRT and MAGIC data do not show evidence for a significant periodicity, as the p-values suggest that the data behaviour is fully consistent with the null hypothesis. The \textit{Swift}-UVOT and radio light curves show the strongest \ac{gls} peaks at periods at 806 and 865~days, but cannot be considered significant with global p-values as high as 10\% and 3\%. 

\begin{figure}
    \centering
    \includegraphics[width=0.48\textwidth]{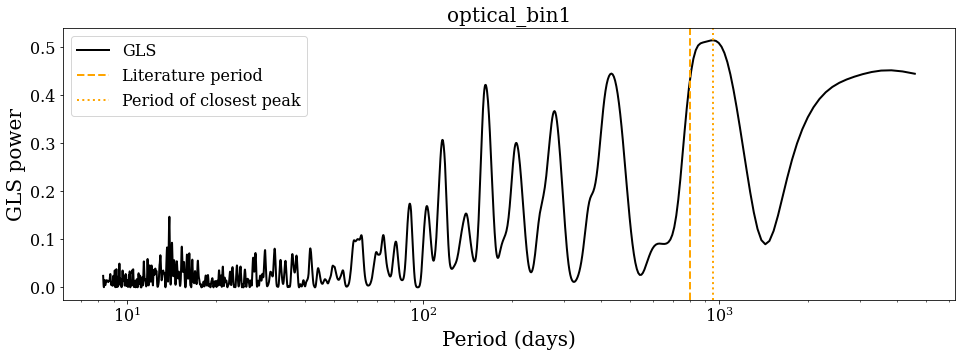}
    \includegraphics[width=0.48\textwidth]{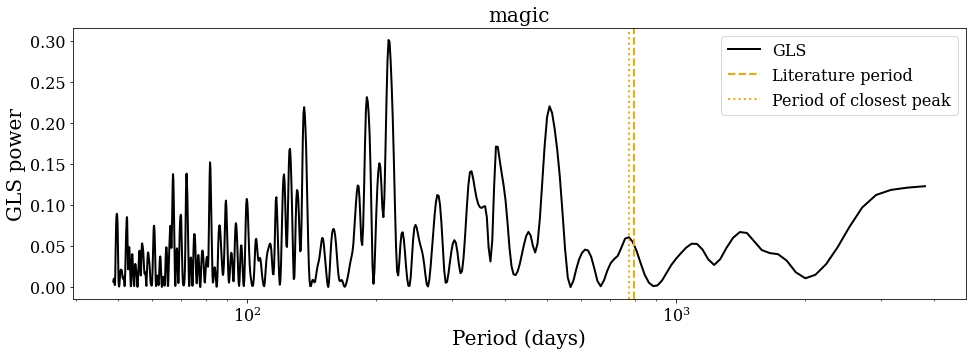}
    \caption{\ac{gls} periodgram of the optical (upper panel) and MAGIC (lower panel) data. The optical data show a broad \ac{gls} peak, described in the text. In case of MAGIC data, no significant peak emerged from the study.}
    \label{fig:gls_opt_magic}
\end{figure}

Before the analysis of the optical light curve, including all the optical data collected in the campaign, we averaged data points within time ranges of one day weighted by the corresponding uncertainties.
In the \ac{gls} of the optical light curve we identify the strongest peak at a period of 957~days with a local p-value of $\sim$10\% and a global p-value of 47\%. This result appears to be in conflict with those of \citet{2018A&A...615A.118S}, who claim a period of $810\pm52$ days with p-values of 1\% or 5\% depending on the method used, and the results of \citet{2021A&A...645A.137A}, who found periods in the range of 801--812~days with uncertainties ranging from 20--70~days, depending on the analysis method and specific band (V or R) with a p-value $<1\%$.
However, we find that the strongest \ac{gls} peak, shown in Fig.~\ref{fig:gls_opt_magic} (upper panel) shows a broad, flat plateau covering the period range from $\sim$800--1000~days. Therefore, the period is poorly constrained. Furthermore, the local p-value strongly depends on the period in that period range. Towards shorter periods the local p-value decreases and is $\sim5\%$ at a period of $\sim800$~days, comparable to one of the results of \citet{2018A&A...615A.118S}. Regarding the optical data we currently do not find convincing evidence for a true periodicity and we conclude that if there truly is a periodicity in the light curve more cycles need to be covered by observations to get a better constraint on the period and the significance.

\begin{table*}
    \centering
    \caption{Results of the search for periodicity in the different bands. }
    \begin{tabular}{lrrrrrr}
        \toprule
        Band & PSD index & Peak period & Peak power & local p-value & global p-value &   p-value \\ 
         & & [d] & & & & (lit. period)\\ 
        \midrule
        Radio 15\,GHz & $2.0\pm0.51$ & 865 & 0.40 & $2.3\times10^{-3}$ & $3.4\times10^{-2}$  & $2.0\times10^{-2}$\\
        Optical & $1.47\pm0.08$ & 957 & 0.51 & $9.7\times10^{-2}$ & $4.7\times10^{-1}$ & $7.8\times10^{-2}$\\ 
        \textit{Swift}-UVOT & $1.41\pm0.12$ & 806 & 0.46 & $5.6\times10^{-3}$ & $1.0\times10^{-1}$ & $5.6\times10^{-3}$\\ 
        \textit{Swift}-XRT & $1.5\pm0.10$ & 2521 & 0.47 & $1.4\times10^{-1}$ & $6.8\times10^{-1}$ & $8.6\times10^{-1}$\\  
        \textit{Fermi}-LAT & $1.4\pm0.26$ & 786 & 0.40 & $2.0\times10^{-5}$ & $1.0\times10^{-3}$ & $2.0\times10^{-5}$\\ 
        MAGIC & $1.0\pm0.42$ & 214 & 0.30 & $1.8\times10^{-2}$ & $3.7\times10^{-1}$ &    $7.0\times10^{-1}$\\ 
        \bottomrule
    \end{tabular} \label{tab:periodicity_results1}
\end{table*}

To complete our analysis, we have estimated the local p-value in the above mentioned \ac{gls} at the literature period of 798 days \citep{2015ApJ...813L..41A}. 
This approach, also in line with the folded lightcurve presented in the previous Section, has the advantage of minimizing the trial factors, since only a single period is tested. The resulting p-values are listed in the last column of Table~\ref{tab:periodicity_results1}.  
The \textit{Fermi}-LAT p-value reflects the local p-value, which is expected as it is the period determined by a subsample of the same data. The radio, optical, \textit{Swift}-UVOT p-values decrease to 2\%, 7.8\%, and 0.56\%, respectively, while the p-values in case of MAGIC and \textit{Swift}-XRT data are well compatible with the null hypothesis with values 70\% and 86\%, respectively. The optical and MAGIC \ac{gls} are diplayed in Figure~\ref{fig:gls_opt_magic}.

\subsection{Further search for periodicity}

A significant indication of periodicity in the gamma-ray light\-curve measured by \textit{Fermi}-LAT was found in several other studies.  Among those, the work by \citet{2020ApJ...896..134P} implemented different methods to detect periodicity in  \textit{Fermi}-LAT blazars. We used the same pipeline described in \citet{2020ApJ...896..134P}, focussing on the Lomb-Scargle and on the wavelet transform approaches, and extending it to the search for periodicity to the VHE, X-ray, UV and optical bands. The details on the two methods can be found in the corresponding paper. 
The results on the {Fermi}-LAT lightcurve is reproduced providing a peak at the period of $\sim 800\,$days with a p-value $< 1\times 10^{-4}$ (pre-trial).
The significance of the periodicity was evaluated with respect to the null hypothesis of a stochastic red noise with the same statistical properties of the lightcurve of the original data. Similar period and significance are obtained on the optical lightcurve, in contrast to the analysis described in the previous section, but slightly compatible with the works \citet[e.g.,][]{2015ApJ...813L..41A,2020ApJ...895..122C,2020ApJ...896..134P}.
As it was commented previously, this can depend on the broad and flat peak in the GLS; any further assessment on the periodicity on the optical band will require further observations. 
The analysis on the X-ray and VHE data confirms the previous findings, showing no significant periodicity (p-value of $>0.1$) at the highest peak.


\section{Discussion and Conclusions}

We have analysed the MWL behaviour of \pg using MAGIC and MWL data from 2007 until the end of 2017 covering bands from radio to VHE gamma rays. The main motivation of this work was to study if the 2.2-year periodicity seen in the GeV gamma rays by the \fermilat\ \citep[][]{2015ApJ...813L..41A} can be seen in our MAGIC data, and if the MWL data can be used to constrain the models explaining the periodicity. 

For these purposes, we have first characterised the variability in the VHE gamma-ray and X-ray bands.
In both cases, we have not found evidence of intraday variability.
Interestingly, intraday variability in the X-ray band was detected by \citet{2017MNRAS.466.3762R} in long \textit{XMM-Newton} observations performed in 2015. From the $\sim 1\,\rm h$ variability time-scale, they inferred a size of the emitting region $R \la \delta \times 10^{14} \, \rm cm$. Recently, \citet{2021MNRAS.506.1198D} confirmed the intraday variability of \pg in the 0.3-10\,keV band with \textit{XMM-Newton} data taken during 2010–2018. The authors found an indication of variability in 16 over 19 observations, where the duration of the observations ranged from 21 to 140\,ks. The doubling timescale ranged from 2 to 33\,ks, i.e. $\sim30$\,min to $\sim$ 9\,h. 
The short duration of our single pointings prevented us to probe intranight variability in $>$hour timescales, as the one suggested in \citet{2021MNRAS.506.1198D} study.

X-ray data in our sample show a hint of harder-when-brighter behaviour often detected in blazars (even if usually it is observed over a shorter timescale). This indicates that possibly freshly injected, high-energy electrons are responsible for the X-ray variability. Furthermore, the long time span considered ensures that the mechanisms driving the spectral variability did not change substantially over time. The same study applied to the MAGIC data gave inconclusive results, probably due to the 1-month averaging applied to the data. A detailed spectral study is planned in a future publication and is beyond the scope of this work.

Interband correlation studies performed on IR, optical, UV, X-ray,  HE and VHE gamma-ray data confirmed the strong IR/optical/UV connection, related to the common synchrotron origin of the emission. An evident correlation between X-rays and VHE gamma rays, and between optical/UV/IR and HE gamma-ray, also emerged from this study, suggesting intertwined emission processes such as that foreseen in the multi-zone, SSC emission scenario. 

Another piece of the puzzle is represented by the detected delayed correlation (of about 3-4 months) between the radio and both optical and \fermilat\ emission \citep{2018MNRAS.480.5517L}. This result is in line with the average behaviour found in gamma-ray detected blazars reported in the same paper. Interestingly, this time-delayed correlation is not present in our radio and X-/VHE gamma-ray data. 

After this detailed characterisation of the variability in general, we have focused on the study of the periodicity. 

A search for evidence of periodicity in the X-ray and TeV bands, as well as in the optical, UV, and GeV bands was performed with a solid statistical approach. 

Our main finding of the periodicity analysis with the \ac{gls} method was that the X-ray and VHE gamma-ray data do not show statistical evidence for a periodic signal. Remarkably, a (hint of a) periodic signal compatible with the one published in literature was found only in the gamma-ray data, which is also the only band with a continous coverage. A solid statistical analysis was applied to the data in the other bands. Radio, UV, and optical data show a periodogram with a peak compatible with the one firmly established in \textit{Fermi}-LAT observations, but with a relatively high p-value (ranging from 4 to $\sim$50\%). 
This is in agreement with the folded MWL lightcurve  built assuming the literature period. The visual inspection of the MWL folded lightcurve suggests a hint of periodic behaviour in the radio, optical, UV, and HE gamma-ray bands. The peak is more pronounced in gamma rays and radio, while it appears broader, and with a similar pattern, in the R and UV2 bands. Polarimetric data, as well as X-ray and VHE gamma-ray data do not show any evident peak in the periodgram. X-ray and VHE gamma-ray folded curves exhibit a similar pattern. Interestingly, a low activity was recorded in all bands at approximately the same phase.


The observed periodicity may be interpreted as a periodic perturbation of the accretion rate on the SMBH and consequently of the fuelling at the base of the jet. The presence of a secondary black hole in a sub-parsec orbit with respect to the primary SMBH originating the jet represents a natural explanation, as previously invoked for OJ 287, despite not unique \citep[e.g.,][]{2018ApJ...866...11D}. Different mechanisms such as jet precession, internal jet rotation, or helical jet motion may also be invoked to explain the periodicity.

The most direct way to constrain a simple precessing jet model would be to observe motion of the jet on the sky. This has been studied in the radio band by \cite{caproni17} who modelled 15\,GHz VLBA data of \pg taken between 2009 and 2016 using a precessing jet model. They modelled the jet using individual Gaussian components, which they then connected to episodes of gamma-ray flares. 
More recently, the radio jet properties of \pg were studied by \cite{2020A&A...634A..87L} using VLBA observations taken between 2015--2017 and lightcurves from OVRO between 2008--2018. While they found clear enhanced activity periods in the radio data, they were not found to be correlated with the 2.2-year gamma-ray periodicity. Moreover, they concluded that the position angle variations of the jet of \pg were not correlated with the gamma-ray periodicity, and a simple geometric model where the variability is caused by changes in the Doppler boosting cannot explain the periodicity, if the gamma-ray and radio variations originate in the same region of the jet. 

\cite{2020A&A...634A..87L} also studied the radio polarisation of \pg using their VLBA data. They found that periods of enhanced polarisation were connected with total intensity flares, indicating that the mechanism producing them is connected. Additionally, they saw a flattening of the radio spectral index at the times of total intensity activity. Such a behaviour could be expected, for example, when shocks compress magnetic field lines, which both increases the fractional polarisation and induces particle acceleration, which flattens the spectral index and increases the total intensity. They also suggested that the low polarisation observed in the core region of \pg is due to multiple polarised components blended within the beam.

Multiple emission components are also supported by the lack of clear periodic modulation in the X-ray and VHE data, which is seen in most of the other wavelengths. On the other hand, the short-term variability in all bands is clearly correlated on some occasions (for example, data around MJD 57000 in Fig.~\ref{fig:corelation_opt-xray_selected} connecting optical and X-ray emission), while at other times there can even be an anti-correlation (data around MJD 57500 in the same Figure). This shows that the situation is very complex. 
The difficulty to connect the low-energy part of the synchrotron bump (IR/optical/UV) to the high-energy synchrotron part (X-rays) was extensively studied in \cite{2015MNRAS.454..353R}. They studied the synchrotron spectrum of the source in multiple activity states and found that the changes in the spectrum can be explained with an inhomogeneous helical jet model, where the high-energy emission originates closer to the black hole than the low-energy emission. Alternatively, there could be multiple (disconnected) emission components or a more complex electron distribution than typically assumed. 

Our analyses on the periodicity show that there clearly must be multiple components contributing to the emission, but that they also cannot be fully disconnected because we (at least sometimes) see simultaneous flaring in all bands. Moreover, the minima in the folded lightcurves seem to be in phase in all bands. Some of the bands (X-ray and VHE) may be more sensitive to the stochastic variations only, while in the other wavelengths, connected with the low-energy part of the SED peaks, we can also see the periodic modulation. This means that any model explaining the periodicity should also be able to explain why it is more prominent in the low-energy part. Confirming this discrepancy would also require longer, densely sampled, lightcurves in X-ray and VHE energies.

\section*{Acknowledgments}{
We would like to thank the Instituto de Astrof\'{\i}sica de Canarias for the excellent working conditions at the Observatorio del Roque de los Muchachos in La Palma. The financial support of the German BMBF, MPG and HGF; the Italian INFN and INAF; the Swiss National Fund SNF; the grants PID2019-104114RB-C31, PID2019-104114RB-C32, PID2019-104114RB-C33, PID2019-105510GB-C31, PID2019-107847RB-C41, PID2019-107847RB-C42, PID2019-107847RB-C44, PID2019-107988GB-C22, PID2020-118491GB-I00 funded by the Spanish MCIN/AEI/ 10.13039/501100011033; the Indian Department of Atomic Energy; the Japanese ICRR, the University of Tokyo, JSPS, and MEXT; the Bulgarian Ministry of Education and Science, National RI Roadmap Project DO1-400/18.12.2020 and the Academy of Finland grant nr. 320045 is gratefully acknowledged. This work was also been supported by Centros de Excelencia ``Severo Ochoa'' y Unidades ``Mar\'{\i}a de Maeztu'' program of the Spanish MCIN/AEI/ 10.13039/501100011033 (SEV-2016-0588, CEX2019-000920-S, CEX2019-000918-M, CEX2021-001131-S, MDM-2015-0509-18-2) and by the CERCA institution of the Generalitat de Catalunya; by the Croatian Science Foundation (HrZZ) Project IP-2016-06-9782 and the University of Rijeka Project uniri-prirod-18-48; by the Deutsche Forschungsgemeinschaft (SFB1491 and SFB876); the Polish Ministry Of Education and Science grant No. 2021/WK/08; and by the Brazilian MCTIC, CNPq and FAPERJ.
E.P. acknowledges funding from Italian Ministry of Education, University and Research (MIUR) through the "Dipartimenti di eccellenza” project Science of the Universe. T.H. was supported by the Academy of Finland projects 317383, 320085, 322535, and 345899.
The Liverpool Telescope is operated on the island of La Palma by Liverpool John Moores University in the Spanish Observatorio del Roque de los Muchachos of the Instituto de Astrofisica de Canarias with financial support from the UK Science and Technology Facilities Council.
Data was used from Steward Observatory blazar spectropolarimetric monitoring project, which was supported by NASA Fermi Guest Investigator Grants NNX09AU10G, NNX12AO93G, and NNX15AU81G.
D.B. acknowledges support from the European Research Council (ERC) under the European Union Horizon 2020 research and innovation program under the grant agreement No 771282.
G.D. and O.V. acknowledge support by the Astronomical station Vidojevica, funding from the Ministry of Science, Technological Development and Innovation of the Republic of Serbia (contract No. 451-03-47/2023-01/200002), by the EC through project BELISSIMA (call FP7-REGPOT-2010-5, No. 265772), the observing and financial grant support from the Institute of Astronomy and Rozhen NAO BAS through the bilateral SANU-BAN joint research project GAIA ASTROMETRY AND FAST VARIABLE ASTRONOMICAL OBJECTS, and support by the SANU project F-187.
The Abastumani team acknowledges financial support by the Shota Rustaveli NSF of Georgia under contract FR-19-6174.
This research was partially supported by the Bulgarian National Science Fund of the Ministry of Education and Science under grants DN 18-13/2017, KP-06-H28/3 (2018), KP-06-H38/4 (2019) and KP-06-KITAJ/2 (2020).
We acknowledge support by Bulgarian National Science Fund under grant DN18-10/2017 and National RI Roadmap Projects D01-383/18.12.2020 of the Ministry of Education and Science of the Republic of Bulgaria.
The research at Boston University was supported by NASA Fermi Guest Investigator Program grants 80NSSC22K1571 and 80NSSC23K1507. This study used observations conducted with the 1.8m Perkins Telescope (PTO) in Arizona (USA), which is owned and operated by Boston University. 
This research has made use of data from the OVRO 40-m monitoring program (Richards, J. L. et al. 2011, ApJS, 194, 29), supported by private funding from the California Institute of Technology and the Max Planck Institute for Radio Astronomy, and by NASA grants NNX08AW31G, NNX11A043G, and NNX14AQ89G and NSF grants AST-0808050 and AST- 1109911. 
W.M. gratefully acknowledges support by the ANID BASAL project FB210003 and FONDECYT 11190853. S.K. acknowledges support from the European Research Council (ERC) under the European Unions Horizon 2020 research and innovation programme under grant agreement No.~771282. R.R. acknowledges support from ANID BASAL projects ACE210002 and FB210003, and ANID Fondecyt 1181620.
R.L. acknowledges financial support from the State Agency for Research of the Spanish MCIU through the "Center of Excellence Severo Ochoa" award to the Instituto de Astrofísica de Andalucía (SEV-2017-0709)
}

\section*{Author Contributions}
E.P., A.S., and T.H. project leadership, paper drafting and edition; J.B.G., P.D.V., L.F., 
S.P., and E.P. MAGIC data analysis, A.S. \textit{Swift} data analysis, I.L. Bayesian block analysis and paper drafting, P.P. and S.K. periodicity analysis and paper drafting. 
C.M.R. and M.V. for the WEBT campaign management and paper drafting.
The rest of the authors have contributed in one or several of the following ways: design, construction, maintenance and operation of the instrument(s) used to acquire the data; preparation and/or evaluation of the observation proposals; data acquisition, processing, calibration and/or reduction; production of analysis tools and/or related Monte Carlo simulations; discussion and approval of the contents of the draft.

\section*{Data Availability}
The data underlying this article will be shared on reasonable request to the corresponding authors.

\bibliographystyle{mnras} 
\bibliography{draft_1553_mwl_10y} 

\section*{Affiliations}
\noindent
{\it
$^{1}$ {Japanese MAGIC Group: Institute for Cosmic Ray Research (ICRR), The University of Tokyo, Kashiwa, 277-8582 Chiba, Japan} \\
$^{2}$ {ETH Z\"urich, CH-8093 Z\"urich, Switzerland} \\
$^{3}$ {Instituto de Astrof\'isica de Canarias and Dpto. de  Astrof\'isica, Universidad de La Laguna, E-38200, La Laguna, Tenerife, Spain} \\
$^{4}$ {Instituto de Astrof\'isica de Andaluc\'ia-CSIC, Glorieta de la Astronom\'ia s/n, 18008, Granada, Spain} \\
$^{5}$ {National Institute for Astrophysics (INAF), I-00136 Rome, Italy} \\
$^{6}$ {Universit\`a di Udine and INFN Trieste, I-33100 Udine, Italy} \\
$^{7}$ {Max-Planck-Institut f\"ur Physik, D-80805 M\"unchen, Germany} \\
$^{8}$ {Universit\`a di Padova and INFN, I-35131 Padova, Italy} \\
$^{9}$ {Institut de F\'isica d'Altes Energies (IFAE), The Barcelona Institute of Science and Technology (BIST), E-08193 Bellaterra (Barcelona), Spain} \\
$^{10}$ {Technische Universit\"at Dortmund, D-44221 Dortmund, Germany} \\
$^{11}$ {Croatian MAGIC Group: University of Zagreb, Faculty of Electrical Engineering and Computing (FER), 10000 Zagreb, Croatia} \\
$^{12}$ {IPARCOS Institute and EMFTEL Department, Universidad Complutense de Madrid, E-28040 Madrid, Spain} \\
$^{13}$ {Centro Brasileiro de Pesquisas F\'isicas (CBPF), 22290-180 URCA, Rio de Janeiro (RJ), Brazil} \\
$^{14}$ {Centro de Investigaciones Energ\'eticas, Medioambientales y Tecnol\'ogicas, E-28040 Madrid, Spain} \\
$^{15}$ {Departament de F\'isica, and CERES-IEEC, Universitat Aut\`onoma de Barcelona, E-08193 Bellaterra, Spain} \\
$^{16}$ {Universit\`a di Pisa and INFN Pisa, I-56126 Pisa, Italy} \\
$^{17}$ {Universitat de Barcelona, ICCUB, IEEC-UB, E-08028 Barcelona, Spain} \\
$^{18}$ {Department for Physics and Technology, University of Bergen, Norway} \\
$^{19}$ {INFN MAGIC Group: INFN Sezione di Catania and Dipartimento di Fisica e Astronomia, University of Catania, I-95123 Catania, Italy} \\
$^{20}$ {INFN MAGIC Group: INFN Sezione di Torino and Universit\`a degli Studi di Torino, I-10125 Torino, Italy} \\
$^{21}$ {INFN MAGIC Group: INFN Sezione di Bari and Dipartimento Interateneo di Fisica dell'Universit\`a e del Politecnico di Bari, I-70125 Bari, Italy} \\
$^{22}$ {Croatian MAGIC Group: University of Rijeka, Faculty of Physics, 51000 Rijeka, Croatia} \\
$^{23}$ {University of Geneva, Chemin d'Ecogia 16, CH-1290 Versoix, Switzerland} \\
$^{24}$ {Japanese MAGIC Group: Physics Program, Graduate School of Advanced Science and Engineering, Hiroshima University, 739-8526 Hiroshima, Japan} \\
$^{25}$ {Armenian MAGIC Group: ICRANet-Armenia, 0019 Yerevan, Armenia} \\
$^{26}$ {University of Lodz, Faculty of Physics and Applied Informatics, Department of Astrophysics, 90-236 Lodz, Poland} \\
$^{27}$ {Croatian MAGIC Group: Josip Juraj Strossmayer University of Osijek, Department of Physics, 31000 Osijek, Croatia} \\
$^{28}$ {Universit\"at W\"urzburg, D-97074 W\"urzburg, Germany} \\
$^{29}$ {Finnish MAGIC Group: Finnish Centre for Astronomy with ESO, University of Turku, FI-20014 Turku, Finland} \\
$^{30}$ {Japanese MAGIC Group: Department of Physics, Tokai University, Hiratsuka, 259-1292 Kanagawa, Japan} \\
$^{31}$ {Universit\`a di Siena and INFN Pisa, I-53100 Siena, Italy} \\
$^{32}$ {Saha Institute of Nuclear Physics, A CI of Homi Bhabha National Institute, Kolkata 700064, West Bengal, India} \\
$^{33}$ {Inst. for Nucl. Research and Nucl. Energy, Bulgarian Academy of Sciences, BG-1784 Sofia, Bulgaria} \\
$^{34}$ {Japanese MAGIC Group: Department of Physics, Yamagata University, Yamagata 990-8560, Japan} \\
$^{35}$ {Finnish MAGIC Group: Space Physics and Astronomy Research Unit, University of Oulu, FI-90014 Oulu, Finland} \\
$^{36}$ {Japanese MAGIC Group: Chiba University, ICEHAP, 263-8522 Chiba, Japan} \\
$^{37}$ {Japanese MAGIC Group: Institute for Space-Earth Environmental Research and Kobayashi-Maskawa Institute for the Origin of Particles and the Universe, Nagoya University, 464-6801 Nagoya, Japan} \\
$^{38}$ {Japanese MAGIC Group: Department of Physics, Kyoto University, 606-8502 Kyoto, Japan} \\
$^{39}$ {INFN MAGIC Group: INFN Sezione di Perugia, I-06123 Perugia, Italy} \\
$^{40}$ {INFN MAGIC Group: INFN Roma Tor Vergata, I-00133 Roma, Italy} \\
$^{41}$ {Japanese MAGIC Group: Department of Physics, Konan University, Kobe, Hyogo 658-8501, Japan} \\
$^{42}$ {also at International Center for Relativistic Astrophysics (ICRA), Rome, Italy} \\
$^{43}$ {now at Institute for Astro- and Particle Physics, University of Innsbruck, A-6020 Innsbruck, Austria} \\
$^{44}$ {also at Port d'Informaci\'o Cient\'ifica (PIC), E-08193 Bellaterra (Barcelona), Spain} \\
$^{45}$ {also at Institute for Astro- and Particle Physics, University of Innsbruck, A-6020 Innsbruck, Austria} \\
$^{46}$ {also at Department of Physics, University of Oslo, Norway} \\
$^{47}$ {also at Dipartimento di Fisica, Universit\`a di Trieste, I-34127 Trieste, Italy} \\
$^{48}$ {Max-Planck-Institut f\"ur Physik, D-80805 M\"unchen, Germany} \\
$^{49}$ {also at INAF Padova} \\
$^{50}$ {Japanese MAGIC Group: Institute for Cosmic Ray Research (ICRR), The University of Tokyo, Kashiwa, 277-8582 Chiba, Japan} \\
$^{51}$ {Astrophysics Research Institute, Liverpool John Moores University and Liverpool Science Park, 146 Brownlow Hill, Liverpool L3 5RF, UK} \\
$^{52}$ {Steward Observatory, University of Arizona, Tucson, AZ 85721, USA} \\
$^{53}$ {Institute of Astrophysics, Foundation for Research and Technology - Hellas, Voutes, 7110 Heraklion, Greece} \\
$^{54}$ {Department of Physics, University of Crete, 71003, Heraklion, Greece} \\
$^{55}$ {INAF, Osservatorio Astrofisico di Torino, via Osservatorio 20, I-10025 Pino Torinese, Italy} \\
$^{56}$ {Ulugh Beg Astronomical Institute, Astronomy Street 33, Tashkent 100052, Uzbekistan}\\
$^{57}$ {Abastumani Observatory, Mt. Kanobili, 0301 Abastumani, Georgia} \\
$^{58}$ {Landessternwarte, Zentrum für Astronomie der Universität Heidelberg, Königstuhl 12, 69117 Heidelberg, Germany} \\
$^{59}$ {EPT Observatories, Tijarafe, E-38780 La Palma, Spain and INAF, TNG Fundación Galileo Galilei, E-38712 La Palma, Spain} \\
$^{60}$ {Saint Petersburg State University, 7/9 Universitetskaya nab., St. Petersburg, 199034 Russia} \\
$^{61}$ {Special Astrophysical Observatory, Russian Academy of Sciences, 369167, Nizhnii Arkhyz, Russia and Pulkovo Observatory, St.Petersburg, 196140, Russia} \\
$^{62}$ {Crimean Astrophysical Observatory RAS, P/O Nauchny, 298409, Crimea} \\
$^{63}$ {Department of Astronomy, Faculty of Physics, University of Sofia, BG-1164 Sofia, Bulgaria} \\
$^{64}$ {Astronomical Observatory, Volgina 7, 11060 Belgrade, Serbia} \\
$^{65}$ {National University of Uzbekistan, Tashkent 100174, Uzbekistan} \\
$^{66}$ {Hans-Haffner-Sternwarte, Naturwissenschaftliches Labor für Schüler am FKG, Friedrich-Koenig-Gymnasium, D-97082 Würzburg, Germany} \\
$^{67}$ {Department of Physics and Astronomy, Faculty of Natural Sciences, University of Shumen, 9712 Shumen, Bulgaria} \\
$^{68}$ {Institute for Astrophysical Research, Boston University, 725 Commonwealth Avenue, Boston, MA 02215, USA} \\
$^{69}$ {Engelhardt Astronomical Observatory, Kazan Federal University, Tatarstan, Russia} \\
$^{70}$ {University of Siena Astronomical Observatory, I-53100, Siena, Italy} \\
$^{71}$ {Institute of Astronomy and National Astronomical Observatory, Bulgarian Academy of Sciences, 1784 Sofia, Bulgaria} \\
$^{72}$ {Department of Physics, University of Colorado Denver, Denver, Colorado  80204, USA} \\ 
$^{73}$ {Owens Valley Radio Observatory, California Institute of Technology,  Pasadena, CA 91125, USA} \\
$^{74}$ {Departamento de Astronomi\'a, Universidad de Chile, Camino El Observatorio 1515, Las Condes, Santiago, Chile}  \\
$^{75}$ {CePIA, Departamento de Astronomía, Universidad de Concepción, Chile}  \\
$^{76}$ {INAF, Osservatorio Astronomico di Brera, Via E. Bianchi 46, I-23807 Merate, Italy} \\
$^{77}$ {Aalto University Mets\"ahovi Radio Observatory, Mets\"ahovintie 114, FI-02540 Kylm\"al\"a, Finland}
$^{78}$ {INAF, Istituto di Radioastronomia, via Gobetti 101, 40129 Bologna, Italy} \\
$^{79}$ {INFN – Sezione di Milano-Bicocca, piazza della Scienza 3, I-20126 Milano (MI), Italy} \\
$^{80}$ {Aryabhatta Research Institute of Observational Sciences (ARIES), Manora Park, Nainital 263 001, India}\\ 
$^{81}$ {NASA Marshall Space Flight Center, Huntsville, AL 35812, USA}\\ 

}

\label{lastpage}
\end{document}